\documentclass{article}

\usepackage{authblk}
\usepackage[font=small]{caption}
\usepackage{amsmath,amssymb,amsfonts}
\usepackage{graphicx}
\usepackage{subfig}
\usepackage{xcolor}
\usepackage[square,numbers]{natbib}
\usepackage{multirow}
\usepackage{soul}

\let\citeleft=(
\let\citeright=)

\begin{document}

\title{Adapting model-based deep learning to multiple acquisition conditions: Ada-MoDL}

\author[1]{Aniket Pramanik}
\author[2]{Sampada Bhave}
\author[2]{Saurav Sajib}
\author[2]{Samir D. Sharma}
\author[1]{Mathews Jacob}

\affil[1]{\small Department of Electrical and Computer Engineering, University of Iowa, Iowa, USA}
\affil[2]{\small Canon Medical Research USA, Inc., Mayfield Village, Ohio, USA}
\maketitle

\clearpage

\section*{Abstract}

\noindent
\textbf{Purpose}: The aim of this work is to introduce a single model-based deep network that can provide high-quality reconstructions from undersampled parallel MRI data acquired with multiple sequences, acquisition settings and field strengths. 

\noindent
\textbf{Methods}: A single unrolled architecture, which offers good reconstructions for multiple acquisition settings, is introduced. The proposed scheme adapts the model to each setting by scaling the CNN features and the regularization parameter with appropriate weights. The scaling weights and regularization parameter are derived using a multi-layer perceptron model from conditional vectors, which represents the specific acquisition setting. The perceptron parameters and the CNN weights are  jointly trained using data from multiple acquisition settings, including differences in field strengths, acceleration, and contrasts. The conditional network is validated using datasets acquired with different acquisition settings. 

\noindent
\textbf{Results}: The comparison of the adaptive framework, which trains a single model using the data from all the settings, shows that it can offer consistently improved performance for each acquisition condition. The comparison of the proposed scheme with networks that are trained independently for each acquisition setting shows that it requires less training data per acquisition setting to offer good performance. 

\noindent
\textbf{Conclusion}: The Ada-MoDL framework enables the use of a single model-based unrolled network for multiple acquisition settings. In addition to eliminating the need to train and store multiple networks for different acquisition settings, this approach reduces the training data needed for each acquisition setting.
  
\noindent
\textbf{Keywords}: Parallel MRI, Unrolled Deep Learning, Adaptive framework, Acquisition setting.

\clearpage

\section*{Introduction}
\label{sec:introduction}
Magnetic Resonance Imaging (MRI) provides an excellent soft tissue contrast at the expense of long acquisition times. Several acceleration methods, including parallel MRI (PMRI) \cite{sodickson1997simultaneous,pruessmann1999sense,griswold2002generalized}, compressed sensing \cite{lustig2007sparse}, and low-rank \cite{nakarmi2017kernel,lingala2011accelerated}, were introduced to speed up the acquisition. The reconstruction algorithms often pose the image recovery from undersampled measurements as a regularized optimization scheme. In recent years, deep-learning (DL) based algorithms have shown immense power in learning the approximate distribution/manifold of images, giving improved reconstruction performance over the non-DL methods discussed above. These methods include direct inversion schemes and model-based strategies. Direct inversion schemes use a CNN (for example, UNET, ResNet) to map/invert an undersampled image to a fully sampled one \cite{han2019k, ghodrati2019mr}. The mapping does not incorporate knowledge about data acquisition physics. Model-based algorithms integrate a CNN-based learned model with a physical model encoding the data acquisition process to obtain a solution consistent with the measurements \cite{lecun2015deep,hammernik2018learning,aggarwal2018modl,pramanik2020deep, hammernik2022machine, hammernik2022physics, adler2018learned, hammernik2020sigma, liang2020deep}. These algorithms are trained using algorithm unrolling; an iterative algorithm is unrolled assuming a finite number of iterations, followed by an end-to-end training. Such schemes have shown benefit in leveraging physics-based acquisition models to obtain improved performance, compared to direct inversion schemes. These algorithms often require fewer training datasets \cite{aggarwal2018modl} compared to direct inversion schemes and also have the ability to incorporate multiple blocks of priors for regularizing the solution. These model-based algorithms offer improved performance because the learned CNN representation is closely linked to the acquisition settings, image contrast, and signal-to-noise ratio (SNR). 

We note that in a clinical setup, a typical MRI exam often consists of several different sequences, each with different acquisition settings that differ in SNR, acceleration, and contrast. The image content also significantly differs depending on the anatomy, the acquisition settings chosen for the specific contrast, and field strength. Most of the above DL schemes may be termed as conditional models because they are trained with paired undersampled and fully sampled acquisitions. A challenge with such conditional schemes, which may restrict their clinical deployment, is the dependence of the learned representation on the specific acquisition setting. It is well-known that changes in the acquisition scheme from the one the network is trained for can result in degradation in performance \cite{gilton2021model}. Similar findings have also been reported in \cite{dar2022adaptive}, where conditional networks have shown degradation with domain shifts. While the training procedure can be modified to include datasets obtained through multiple acquisition operators to reduce the sensitivity of the learned representation to the acquisition scheme \cite{aggarwal2018modl}, this approach often comes at the expense of reduced performance in each specific setting. Thus, it becomes necessary to train different models for different anatomy or acquisition settings that are optimal for a particular setting.

In order to provide optimum reconstruction performance, conditional models would need to be trained separately for each kind of acquisition. In case of model-based DL schemes, the regularization parameter also needs to be adapted based on the SNR and undersampling rate. The large number of possible acquisition settings makes it challenging to train a model for each of them, especially because a sufficient number of fully sampled training datasets should be acquired in those settings \cite{willemink2020preparing, lundervold2019overview}. In addition, storing multiple networks for deployment would require memory, and switching between the models depending on subtle changes in the acquisition settings can be inconvenient.    

To overcome the above challenges, we introduce a conditional unrolled architecture, termed as Ada-MoDL; the proposed scheme is the conditional extension of the unrolled MoDL framework \cite{aggarwal2018modl}. We use a conditional vector that specifies the acquisition setting (e.g. image contrast, field-strength). Similar to traditional unrolling methods, the proposed scheme also alternates between data-consistency and CNN denoising blocks. The main distinction is that we modulate the output of each CNN layer by a set of scalar weights, which are dependent on the conditional vectors. Similarly, the regularization parameter $\lambda$ that balances the data-consistency and denoising penalties in model-based schemes \cite{aggarwal2018modl} is also made dependent on the conditional vectors. The modulation enables the selection of features generated by the CNN, based on acquisition information. The dependence between the scalar weights  and the conditional vectors are modeled by a multi-layer perceptron \cite{jain1996artificial} (MLP) model that maps a conditional vector to a vector of scaling factors and the regularization parameter. In addition to weights for the CNN features, the regularization parameter in the data-consistency block is also learned as a function of the conditional vectors by the MLP. Similar to the MoDL framework that shares the CNN weights across iterations, the feature scaling weights learned by MLP are also shared across iterations. This work is inspired by adaptive instance normalization schemes (Ada-IN) that are widely used in deep learning \cite{huang2017arbitrary, elmas2022federated, dalmaz2022one}. Recent works \cite{dar2022adaptive, korkmaz2022unsupervised} use subject-specific re-training of the learned model during inference to account for each acquisition setting, which is computationally expensive. By contrast, Ada-MoDL learns a model for a pre-determined set of acquisition conditions, which offers fast inference, similar to unrolled algorithms. 

The number of free parameters in the MLP is approximately 5\% of the number of parameters of the CNN; the proposed extension involves minimal overhead. This architecture enables the adaptation of the network to different acquisition settings, which allows one to train the network using data from different acquisition settings. In this work, we account for differences in contrast, acceleration factor, and field strength. The joint training strategy enables the combination of information from multiple acquisition settings and hence is expected to be more training data efficient than the individual training of unrolled networks for each acquisition setting.

\section*{Theory}
\label{sec:theory}

We consider recovery of an image $\mathbf x \in \mathbb{C^N}$ from its set of measurements obtained through a parallel MRI acquisition satisfying,
\begin{equation}
\label{ill-posed}
\mathbf b = \mathcal A(\mathbf x) + \mathbf n 
\end{equation}
where $\mathcal A$ is a linear operator embedding point-wise multiplication with coil sensitivity maps, the Fourier transform operator, and an undersampling operator. The vector $\mathbf b$ represents a set of noisy measurements, while $\mathbf n$ denotes additive Gaussian noise. 

\subsection*{Brief review of current inversion methods}
A common practice to reduce scan time is to undersample the multi-channel Fourier space. The measurement operator in this case is often poorly conditioned, or even low-rank. 
\subsubsection*{Classical inversion methods}
Several regularized inversion strategies have been introduced to make the recovery well-conditioned. These schemes pose the recovery as an optimization problem,
\begin{equation}
\label{regularized_recovery}
\arg \min_{\mathbf x} \frac{\lambda}{2}\|\mathcal A(\mathbf x) - \mathbf b\|_2^2 + \mathcal R(\mathbf x)
\end{equation}
where $\mathcal R(\mathbf x)$ is a regularization term constraining the solution by using prior information about $\mathbf x$. The scalar $\lambda$ is a tunable parameter balancing the effect of data consistency (DC) and prior term. Traditional regularizers include handcrafted priors such as $l_1$ norm of wavelets \cite{lustig2007sparse}, total variation \cite{block2007undersampled, knoll2011second}, low-rank \cite{nakarmi2017kernel,lingala2011accelerated}, structured low-rank methods \cite{lee2016acceleration,haldar2016p,shin2014calibrationless,ongie2017fast}, and sparsity in other transform domains \cite{akccakaya2011low, ravishankar2010mr, doneva2010compressed}. 

Plug and play (PnP) models rely on off-the-shelf or pre-trained CNN denoisers to offer improved reconstruction. These methods consider an iterative proximal gradient algorithm used for sparse recovery; the proximal mapping steps in these algorithms are often replaced by denoisers to obtain improved results as compared to regularized inversion using classical penalties \cite{venkatakrishnan2013plug}. The above regularization schemes including PnP models require the tuning of the regularization parameter $\lambda$ to obtain good reconstructions. 

\subsubsection*{Deep unrolled networks}

Recently, several authors have introduced deep unrolled algorithms, which offer further improvement in performance  \cite{ongie2020deep, hammernik2018learning,aggarwal2018modl,sun2016deep}. Unrolled algorithms assume a specific $\mathcal A$ operator and unroll the above iterative proximal gradients algorithm, assuming a finite number of iterations. The resulting deep network, which alternates between DC blocks and DL blocks, is trained in an end-to-end fashion, such that the reconstructed image matches the original image. The regularization parameter $\lambda$ is also optimized during training. 
Deep unrolled methods train the CNN blocks such that the reconstructed image best matches the fully sampled image \cite{hammernik2018learning,xiang2021fista,aggarwal2018modl,sun2016deep}. In this work, we focus on the MoDL algorithm, which was previously introduced for parallel MRI  \cite{aggarwal2018modl}. We note that the proposed framework is broadly applicable to general unrolled algorithms including \cite{hammernik2018learning,sun2016deep,xiang2021fista} and direct inversion \cite{han2019k} methods. MoDL replaces the prior term $\mathcal R(\mathbf x)$ with a residual CNN that learns noise and aliasing patterns in the image to denoise it:
\begin{equation}
\label{modl}
\mathbf x = \arg \min_{\mathbf x} \frac{\lambda}{2}\|\mathcal A(\mathbf x) - \mathbf b\|_2^2 + \|\mathcal N(\mathbf x, \theta)\|_2^2
\end{equation}
where $\mathcal N(\mathbf x, \theta) = \mathcal I(\mathbf x) - \mathcal D(\mathbf x, \theta)$ is a CNN. The MoDL architecture is shown in Fig \ref{fig:arch}(a). Here, $\theta$ denotes trainable CNN weights. $\mathcal N(\mathbf x, \theta)$ learns noise/alias patterns from the input and subtracts those from the image itself to provide a denoised image $\mathcal D(\mathbf x, \theta)$. The scalar $\lambda$ is a trainable parameter that controls the trade-off between DC and denoising operation. 

MoDL solves the optimization problem in \eqref{modl} by alternating between the following steps:
\begin{eqnarray}
\label{cnn}
\mathbf z_k &=& \mathcal D(\mathbf x_k, \theta) \\ 
\label{dc}
\mathbf x_{k+1} &=& (\mathcal A^H\mathcal A + \lambda \mathcal I)^{-1}(\mathcal A^H\mathbf b + \lambda \mathbf z_k).
\end{eqnarray}
Here, \eqref{cnn} denotes the denoising operation on updated image $\mathbf x$  and \eqref{dc} is the DC step performing physics-based image update on the denoised output $\mathbf z$.

\subsection*{Challenges with deep unrolled methods}
Empirical results show that the unrolled optimization offers improved performance when compared to model-agnostic PnP methods \cite{ongie2020deep, hammernik2018learning,aggarwal2018modl,sun2016deep,wang2016accelerating,xiang2021fista}, because it learns a representation that is ideal for a specific measurement operator. However, the performance improvement often comes at the expense of reduced generalizability to measurement conditions \cite{gilton2021model}. In particular, the unrolled model trained for a specific $\mathcal A$ may result in sub-optimal results for another measurement scheme (e.g., a different undersampling rate or measurement condition). As shown in \cite{aggarwal2018modl}, the generalizability may be improved by training with multiple measurement models, at the expense of reduced performance. 

MRI offers improved visualization of tissue using multiple contrasts (e.g., $T_1$, $T_2$, FLAIR). Each of these measurement schemes uses very different acquisition settings. For instance, the SNR of inversion-recovery based FLAIR acquisitions is often lower than that of $T_1$ or $T_2$ scans. Since FLAIR often requires a higher scan time, it is a common practice to choose a higher undersampling rate for FLAIR, which makes recovery even more challenging. Similarly, the SNR of images varies with scanner field strengths. In many cases, specific acceleration factors are chosen to match the demands of the scan. 

When unrolled architectures are used, one has to train multiple CNN modules for each of the above acquisition setups to get the best performance. The acquisition of a large number of fully sampled exemplar datasets for each of the settings is often challenging. In addition,  storage and deployment of models for each of the different contrasts, field strengths, and acceleration factors is also required. While one may train a single model for different acquisition settings, this approach often translates to degraded performance as discussed earlier.

\subsection*{Ada-MoDL}
We propose to overcome these limitations by introducing a conditional MoDL architecture, where a single model is adapted depending on the acquisition settings. Because we expect the network parameters to vary with image contrast, SNR, and other acquisition settings, we propose to use these parameters as conditional vector denoted by $m$; the conditional vector may be derived from the meta-data of the acquisition. 

\begin{figure*}[t!]
\centering
\includegraphics[width=\textwidth,keepaspectratio=true,trim={1.7cm 6cm 1.7cm 6.3cm},clip]{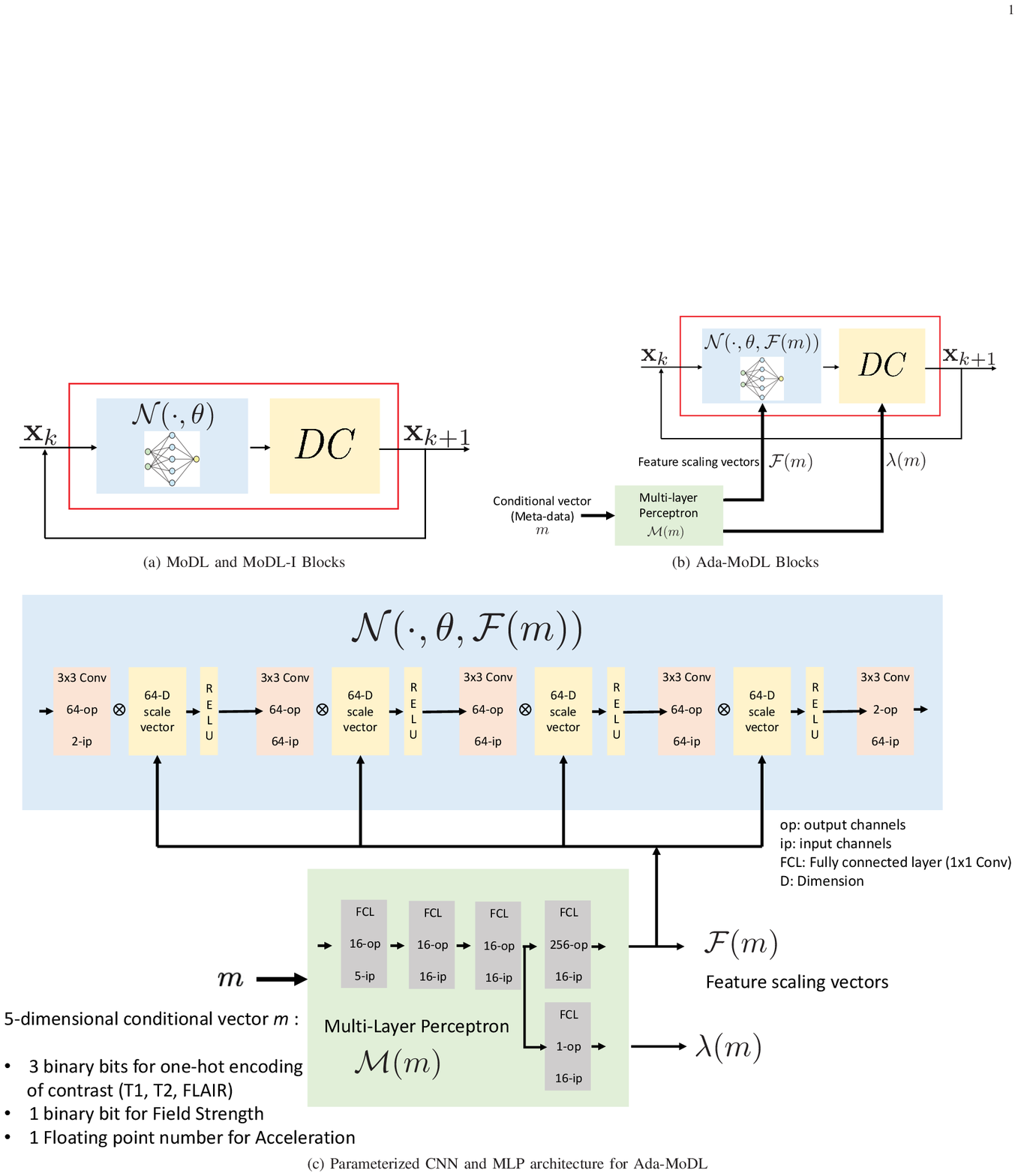}
\caption{An example architecture for the MoDL, MoDL-I, and Ada-MoDL framework, which we use for the first dataset. The figure (a) shows recursive network for MoDL and MoDL-I, which consists of a five-layer CNN $\mathcal N(\cdot, \theta)$ from \eqref{cnn} and a DC block \eqref{dc}; (b) shows the recursive architecture for Ada-MoDL where MLP $\mathcal M(m)$ maps the acquisition parameters represented by the conditional vector $m$ of the dataset to feature scaling weights  $\mathcal F(m)$ and regularization parameter $\lambda(m)$ in \eqref{meta_modl}. $\mathcal F(m)$ consists of feature-scaling vectors for the output features of the four intermediate CNN layers $\mathcal N(\cdot, \theta, \mathcal F(m))$. Here, ip denotes the number of input channels and op denotes the number of output channels. FCL denotes a fully connected layer. Ada-MoDL alternates between CNN $\mathcal N(\cdot, \theta, \mathcal F(m))$ from \eqref{cnn_meta} and a DC block \eqref{dc_meta}. The figure (c) shows a more detailed architecture of $\mathcal N(\cdot, \theta, \mathcal F(m))$ and $\mathcal M(m)$. We use a similar architecture with a UNET denoising network and a similar MLP with a matching number of output channels for the fastMRI dataset.}
\label{fig:arch}
\end{figure*}

Similar to the traditional MoDL, the proposed \emph{denoising} module consists of multiple convolution layers. The key difference with MoDL is that we scale each of the features of the CNN by appropriate weights, which are derived from the conditional vector as $\mathcal F(m)$. Here, $\mathcal F$ is a function that is realized by an MLP.  The modulation of the features allows the approach to emphasize or de-emphasize specific features, depending on the condition vectors. This work is motivated by adaptive instance normalization schemes that are used in computer vision \cite{huang2017arbitrary}. We also propose to express the regularization parameter $\lambda$ as a function of the conditional variables $m$. This adaptation of the regularization parameters allows the network to account for differences in SNR in the datasets. The Ada-MoDL formulation can be compactly represented as,

\begin{equation}
\label{meta_modl}
\mathbf x = \arg \min_{\mathbf x} \frac{\lambda(m)}{2}\|\mathcal A(\mathbf x) - \mathbf b\|_2^2 + \|\mathcal N(\mathbf x, \theta, \mathcal F(m))\|_2^2
\end{equation}
where $\mathcal M(m) = [\mathcal F(m), \lambda(m)]$ is an MLP mapping the conditional vector $m$ to an appropriate vector $\mathcal F(m)$ containing feature scaling factors for CNN $\mathcal N$ and the regularization parameter $\lambda(m)$ for the dataset. Similar to MoDL \cite{aggarwal2018modl}, both CNN and MLP parameters are repeated across iterations. The network is unrolled for a fixed number of iterations, alternating between the DC enforcing step in \eqref{dc_meta} and the residual CNN denoiser in \eqref{cnn_meta}, as shown in Fig. \ref{fig:arch}(b). The equations for alternating blocks are,
\begin{eqnarray}
\label{cnn_meta}
\mathbf z_k &=& \mathcal D(\mathbf x_k, \theta, \mathcal F(m)) \\ 
\label{dc_meta}
\mathbf x_{k+1} &=& (\mathcal A^H\mathcal A + \lambda(m) \mathcal I)^{-1}(\mathcal A^H\mathbf b + \lambda(m) \mathbf z_k),\\\nonumber&&\qquad k=1,..,K
\end{eqnarray}
which provide conditional-vector dependent denoised output $\mathbf z$ and updated image $\mathbf x$. Here, $K$ is the number of unrolls or iterations. Note that the same denoiser block $\mathcal D$ is used for all the iterations $k = 1,..,K$.

An example Ada-MoDL framework with its CNN and MLP architectures is shown in Fig. \ref{fig:arch}(b) and \ref{fig:arch}(c), respectively. The CNN $\mathcal N$ consists of five 3 x 3 convolution layers with 64 output channels for each of them except the last layer (which contains two). The 64 output features from the intermediate layers are scaled by a corresponding 64-dimensional vector obtained from the MLP $\mathcal M$, which is followed by ReLU non-linearity. The MLP $\mathcal M$ consists of five fully connected (1 x 1 convolution) layers with 16 output channels for the intermediate layers. The output layer of $\mathcal M$ provides a vector $\mathcal F(m)$ of size 256 (4 x 64 features per intermediate layer) and a scalar $\lambda(m)$. We use this framework to train a joint model to recover $T_1$, $T_2$, FLAIR datasets with multiple accelerations and two different field strengths. The conditional vector $m$ in this case is five-dimensional, corresponding to each of the different settings. This general strategy can be adapted to other unrolled architectures; we used a similar UNET based model for fastMRI recovery as described in the experiment section.

\begin{table}[h!]
\fontsize{8}{7}
\selectfont
\centering
\begin{tabular}{|c|cc|cc|cc|cc|}
\cline{2-9}
\multicolumn{1}{c|}{} & \multicolumn{8}{|c|}{FLAIR-3T} \\ \cline{2-9}
\multicolumn{1}{c|}{} & \multicolumn{2}{c|}{4.0x} & \multicolumn{2}{c|}{3.5x} & \multicolumn{2}{c|}{2.5x} & \multicolumn{2}{c|}{1.8x}\\  
\multicolumn{1}{c|}{} & PSNR & SSIM & PSNR & SSIM & PSNR & SSIM & PSNR & SSIM \\ \hline 
MoDL & 42.0 $\pm$ 2.1 & 992 $\pm$ 4 & 42.9 $\pm$ 1.9 & 993 $\pm$ 3 & 45.1 $\pm$ 2.3 & 996 $\pm$ 3 & 46.0 $\pm$ 1.4 & 997 $\pm$ 2 \\
Ada-MoDL & \textbf{42.9 $\pm$ 1.7} & \textbf{993 $\pm$ 3} & \textbf{43.6 $\pm$ 2.0} & \textbf{994 $\pm$ 3} & \textbf{46.0 $\pm$ 1.5} & \textbf{997 $\pm$ 1} & 46.6 $\pm$ 1.9 & 997 $\pm$ 2 \\
MoDL-I & 41.6 $\pm$ 2.5 & 992 $\pm$ 4 & 43.1 $\pm$ 1.8 & 993 $\pm$ 4 & 45.9 $\pm$ 2.4 & 997 $\pm$ 1 & \textbf{46.8 $\pm$ 1.5} & \textbf{998 $\pm$ 1}  \\ \hline
\multicolumn{1}{c|}{} & \multicolumn{8}{|c|}{$T_2$-3T} \\ \cline{2-9}
\multicolumn{1}{c|}{} & \multicolumn{2}{c|}{4.0x} & \multicolumn{2}{c|}{3.5x} & \multicolumn{2}{c|}{2.5x} & \multicolumn{2}{c|}{1.8x}\\  
\multicolumn{1}{c|}{} & PSNR & SSIM & PSNR & SSIM & PSNR & SSIM & PSNR & SSIM \\ \hline 
MoDL & 42.3 $\pm$ 1.9 & 992 $\pm$ 5 & 43.2 $\pm$ 2.0 & 993 $\pm$ 4 & 45.6 $\pm$ 1.7 & 996 $\pm$ 3 & 46.6 $\pm$ 1.8 & 997 $\pm$ 2 \\
Ada-MoDL & \textbf{43.5 $\pm$ 1.6} & \textbf{993 $\pm$ 3} & \textbf{44.0 $\pm$ 1.7} & \textbf{994 $\pm$ 4} & \textbf{46.4 $\pm$ 1.4} & \textbf{997 $\pm$ 1} & 47.3 $\pm$ 2.1 & 998 $\pm$ 1  \\
MoDL-I & 42.6 $\pm$ 2.1 & 992 $\pm$ 5 & 43.7 $\pm$ 1.8 & 993 $\pm$ 4 & 46.3 $\pm$ 1.6 & 997 $\pm$ 2 & \textbf{47.5 $\pm$ 1.9} & \textbf{998 $\pm$ 1}  \\ \hline
\multicolumn{1}{c|}{} & \multicolumn{8}{|c|}{FLAIR-1.5T} \\ \cline{2-9}
\multicolumn{1}{c|}{} & \multicolumn{2}{c|}{4.0x} & \multicolumn{2}{c|}{3.5x} & \multicolumn{2}{c|}{2.5x} & \multicolumn{2}{c|}{1.8x}\\  
\multicolumn{1}{c|}{} & PSNR & SSIM & PSNR & SSIM & PSNR & SSIM & PSNR & SSIM \\ \hline 
MoDL & 37.2 $\pm$ 2.2 & 970 $\pm$ 6 & 38.4 $\pm$ 2.0 & 972 $\pm$ 5 & 38.8 $\pm$ 1.8 & 973 $\pm$ 6 & 40.3 $\pm$ 1.9 & 977 $\pm$ 7 \\
Ada-MoDL & \textbf{37.9 $\pm$ 1.9} & \textbf{972 $\pm$ 5} & \textbf{39.6 $\pm$ 2.0} & \textbf{975 $\pm$ 6} & \textbf{40.0 $\pm$ 1.7} & \textbf{978 $\pm$ 4} & \textbf{41.5 $\pm$ 2.1} & \textbf{981 $\pm$ 6}  \\
MoDL-I & 36.9 $\pm$ 2.2 & 970 $\pm$ 6 & 39.4 $\pm$ 2.4 & 975 $\pm$ 7 & 40.0 $\pm$ 1.8 & 978 $\pm$ 6 & 41.1 $\pm$ 2.0 & 981 $\pm$ 5 \\ \hline
\multicolumn{1}{c|}{} & \multicolumn{8}{|c|}{$T_2$-1.5T} \\ \cline{2-9}
\multicolumn{1}{c|}{} & \multicolumn{2}{c|}{4.0x} & \multicolumn{2}{c|}{3.5x} & \multicolumn{2}{c|}{2.5x} & \multicolumn{2}{c|}{1.8x}\\  
\multicolumn{1}{c|}{} & PSNR & SSIM & PSNR & SSIM & PSNR & SSIM & PSNR & SSIM \\ \hline 
MoDL & 38.6 $\pm$ 1.9 & 977 $\pm$ 5 & 39.4 $\pm$ 1.4 & 981 $\pm$ 6 & 40.7 $\pm$ 2.1 & 984 $\pm$ 7 & 42.4 $\pm$ 2.1 & 987 $\pm$ 5 \\
Ada-MoDL & \textbf{39.3 $\pm$ 1.3} & \textbf{979 $\pm$ 6} & \textbf{40.5 $\pm$ 1.7} & \textbf{984 $\pm$ 6} & 41.5 $\pm$ 2.1 & 985 $\pm$ 6 & \textbf{43.0 $\pm$ 2.2} & \textbf{988 $\pm$ 5}  \\
MoDL-I & 38.4 $\pm$ 1.8 & 977 $\pm$ 6 & 40.2 $\pm$ 2.3 & 983 $\pm$ 5 & \textbf{41.6 $\pm$ 1.8} & \textbf{985 $\pm$ 5} & 42.8 $\pm$ 1.6 & 988 $\pm$ 6 \\ \hline
\multicolumn{1}{c|}{} & \multicolumn{8}{|c|}{$T_1$-1.5T} \\ \cline{2-9}
\multicolumn{1}{c|}{} & \multicolumn{2}{c|}{4.0x} & \multicolumn{2}{c|}{3.5x} & \multicolumn{2}{c|}{2.5x} & \multicolumn{2}{c|}{1.8x}\\  
\multicolumn{1}{c|}{} & PSNR & SSIM & PSNR & SSIM & PSNR & SSIM & PSNR & SSIM \\ \hline 
MoDL & 41.6 $\pm$ 2.4 & 991 $\pm$ 5 & 42.2 $\pm$ 1.8 & 992 $\pm$ 5 & 43.8 $\pm$ 2.0 & 995 $\pm$ 3 & 44.6 $\pm$ 2.2 & 997 $\pm$ 1 \\
Ada-MoDL & \textbf{42.5 $\pm$ 2.3} & \textbf{992 $\pm$ 5} & \textbf{43.3 $\pm$ 1.9} & \textbf{994 $\pm$ 3} & \textbf{44.6 $\pm$ 1.7} & \textbf{997 $\pm$ 2} & \textbf{45.1 $\pm$ 2.1} & \textbf{998 $\pm$ 1}  \\
MoDL-I & 41.8 $\pm$ 1.8 & 991 $\pm$ 7 & 42.9 $\pm$ 2.1 & 993 $\pm$ 5 & 44.5 $\pm$ 1.6 & 997 $\pm$ 1 & 44.8 $\pm$ 2.0 & 997 $\pm$ 2  \\ \hline
\end{tabular}
\vspace{1em}
\caption{Performance comparison of Ada-MoDL against MoDL and MoDL-I on the in-house brain dataset. Each sub-table corresponds to a specific acquisition setting. The PSNR and SSIM results for each setting are averaged over six subjects from each setting. The models have been trained and tested on datasets from five different acquisition settings (FLAIR-3T, $T_2$-3T, FLAIR-1.5T, $T_2$-1.5T, and $T_1$-1.5T). The MoDL-I is trained for a specific acquisition setting with only one subject from that setting, while Ada-MoDL and MoDL are trained jointly on the data from all acquisition settings (one subject per setting). }
\label{tab:canon_comp_tab} 
\end{table}

\begin{table}[h!]
\fontsize{8}{7}
\selectfont
\centering
\begin{tabular}{|c|cc|cc|cc|cc|}
\hline
\multicolumn{9}{c}{\multirow{2}{*}{Performance Comparisons on Within-Domain Data}} \\  
\multicolumn{9}{c}{\multirow{2}{*}{}} \\ \hline
\multicolumn{1}{c|}{} & \multicolumn{8}{|c|}{$T_1$} \\ \cline{2-9}
\multicolumn{1}{c|}{} & \multicolumn{2}{c|}{4.0x} & \multicolumn{2}{c|}{3.5x} & \multicolumn{2}{c|}{2.5x} & \multicolumn{2}{c|}{1.8x}\\  
\multicolumn{1}{c|}{} & PSNR & SSIM & PSNR & SSIM & PSNR & SSIM & PSNR & SSIM \\ \hline 
MoDL & 39.1 $\pm$ 2.5 & 990 $\pm$ 6  & 39.9 $\pm$ 2.3 & 991 $\pm$ 5 & 40.9 $\pm$ 2.1 & 993 $\pm$ 4 & 42.7 $\pm$ 2.0 & 995 $\pm$ 3 \\
Ada-MoDL & \textbf{40.4 $\pm$ 1.9} & \textbf{992 $\pm$ 4} & \textbf{40.6 $\pm$ 2.2} & \textbf{993 $\pm$ 4} & \textbf{41.9 $\pm$ 2.0} & \textbf{994 $\pm$ 3} & \textbf{43.4 $\pm$ 1.7} & \textbf{996 $\pm$ 2}  \\
MoDL-I-6 & 39.0 $\pm$ 2.4 & 990 $\pm$ 6 & 40.0 $\pm$ 2.5 & 991 $\pm$ 5 & 41.0 $\pm$ 2.3 & 993 $\pm$ 5 & 42.6 $\pm$ 2.3 & 995 $\pm$ 2  \\
MoDL-I-100 & 39.9 $\pm$ 1.9 & 991 $\pm$ 4 & 40.4 $\pm$ 2.6 & 992 $\pm$ 4 & 41.8 $\pm$ 2.4 & 994 $\pm$ 3 & 43.2 $\pm$ 2.8 & 996 $\pm$ 2  \\ \hline
\multicolumn{1}{c|}{} & \multicolumn{8}{|c|}{$T_2$} \\ \cline{2-9}
\multicolumn{1}{c|}{} & \multicolumn{2}{c|}{4.0x} & \multicolumn{2}{c|}{3.5x} & \multicolumn{2}{c|}{2.5x} & \multicolumn{2}{c|}{1.8x}\\  
\multicolumn{1}{c|}{} & PSNR & SSIM & PSNR & SSIM & PSNR & SSIM & PSNR & SSIM \\ \hline 
MoDL & 38.8 $\pm$ 2.3 & 986 $\pm$ 5 & 39.7 $\pm$ 2.6 & 989 $\pm$ 7 & 42.2 $\pm$ 2.1 & 993 $\pm$ 3 & 42.8 $\pm$ 2.2 & 994 $\pm$ 2 \\
Ada-MoDL & \textbf{39.9 $\pm$ 2.0} & \textbf{989 $\pm$ 6} & \textbf{41.4 $\pm$ 2.4} & \textbf{992 $\pm$ 4} & \textbf{43.1 $\pm$ 2.5} & \textbf{995 $\pm$ 2} & \textbf{43.9 $\pm$ 1.9} & \textbf{996 $\pm$ 2}  \\
MoDL-I-6 & 38.3 $\pm$ 2.2 & 985 $\pm$ 6 & 39.3 $\pm$ 2.0 & 989 $\pm$ 5 & 41.8 $\pm$ 2.5 & 992 $\pm$ 4 & 42.3 $\pm$ 1.8 & 993 $\pm$ 3  \\
MoDL-I-100 & 39.6 $\pm$ 2.7 & 989 $\pm$ 5 & 41.1 $\pm$ 2.6 & 992 $\pm$ 4 & 42.8 $\pm$ 2.3 & 994 $\pm$ 3 & 43.6 $\pm$ 2.4 & 996 $\pm$ 2  \\ \hline
\multicolumn{1}{c|}{} & \multicolumn{8}{|c|}{FLAIR} \\ \cline{2-9}
\multicolumn{1}{c|}{} & \multicolumn{2}{c|}{4.0x} & \multicolumn{2}{c|}{3.5x} & \multicolumn{2}{c|}{2.5x} & \multicolumn{2}{c|}{1.8x}\\  
\multicolumn{1}{c|}{} & PSNR & SSIM & PSNR & SSIM & PSNR & SSIM & PSNR & SSIM \\ \hline 
MoDL & 36.0 $\pm$ 1.9 & 980 $\pm$ 4 & 36.9 $\pm$ 1.8 & 982 $\pm$ 6 & 39.2 $\pm$ 2.0 & 987 $\pm$ 6 & 41.0 $\pm$ 2.0 & 990 $\pm$ 7 \\
Ada-MoDL & \textbf{37.3 $\pm$ 2.1} & \textbf{983 $\pm$ 5} & \textbf{38.2 $\pm$ 1.7} & \textbf{985 $\pm$ 4} & \textbf{40.1 $\pm$ 2.0} & \textbf{989 $\pm$ 6} & \textbf{42.1 $\pm$ 2.2} & \textbf{992 $\pm$ 5}  \\
MoDL-I-6 & 35.6 $\pm$ 2.5 & 980 $\pm$ 5 & 36.2 $\pm$ 2.0 & 980 $\pm$ 5 & 39.0 $\pm$ 2.3 & 987 $\pm$ 6 & 40.7 $\pm$ 2.1 & 990 $\pm$ 4 \\
MoDL-I-100 & 36.9 $\pm$ 2.6 & 982 $\pm$ 7 & 37.7 $\pm$ 2.3 & 985 $\pm$ 5 & 39.5 $\pm$ 1.9 & 988 $\pm$ 4 & 41.6 $\pm$ 2.2 & 991 $\pm$ 5 \\ \hline
\multicolumn{1}{c|}{} & \multicolumn{8}{|c|}{$T_1$-POST} \\ \cline{2-9}
\multicolumn{1}{c|}{} & \multicolumn{2}{c|}{4.0x} & \multicolumn{2}{c|}{3.5x} & \multicolumn{2}{c|}{2.5x} & \multicolumn{2}{c|}{1.8x}\\  
\multicolumn{1}{c|}{} & PSNR & SSIM & PSNR & SSIM & PSNR & SSIM & PSNR & SSIM \\ \hline 
MoDL & 42.2 $\pm$ 1.9 & 994 $\pm$ 3 & 43.0 $\pm$ 2.1 & 995 $\pm$ 2 & 44.8 $\pm$ 2.0 & 997 $\pm$ 1 & 45.6 $\pm$ 2.4 & 998 $\pm$ 1 \\
Ada-MoDL & \textbf{43.1 $\pm$ 2.3} & \textbf{995 $\pm$ 3} & \textbf{43.8 $\pm$ 2.1} & \textbf{996 $\pm$ 2} & \textbf{45.9 $\pm$ 2.2} & \textbf{998 $\pm$ 1} & \textbf{46.3 $\pm$ 1.8} & \textbf{998 $\pm$ 1}  \\
MoDL-I-6 & 42.1 $\pm$ 2.2 & 994 $\pm$ 3 & 43.1 $\pm$ 2.4 & 995 $\pm$ 3 & 45.0 $\pm$ 2.1 & 997 $\pm$ 2 & 45.8 $\pm$ 1.7 & 998 $\pm$ 1 \\
MoDL-I-100 & 43.0 $\pm$ 1.9 & 995 $\pm$ 3 & 43.4 $\pm$ 1.6 & 995 $\pm$ 2 & 45.8 $\pm$ 2.0 & 998 $\pm$ 1 & 46.2 $\pm$ 2.5 & 998 $\pm$ 1 \\ \hline
\multicolumn{9}{c}{\multirow{2}{*}
{Performance Comparisons on Cross-Domain Data}} \\  
\multicolumn{9}{c}{\multirow{2}{*}{}} \\ \hline
\multicolumn{1}{c|}{} & \multicolumn{8}{c|}{Ada-MoDL} \\ \cline{2-9}
\multicolumn{1}{c|}{} & \multicolumn{2}{c|}{$T_1$} & \multicolumn{2}{c|}{$T_2$} & \multicolumn{2}{c|}{FLAIR} & \multicolumn{2}{c|}{$T_1$-POST}\\  
\multicolumn{1}{c|}{} & PSNR & SSIM & PSNR & SSIM & PSNR & SSIM & PSNR & SSIM \\ \hline 
$T_1$ & \textbf{40.4 $\pm$ 1.9} & \textbf{992 $\pm$ 4} & 38.6 $\pm$ 2.3 & 986 $\pm$ 7 & 38.4 $\pm$ 2.5 & 986 $\pm$ 7 & 39.3 $\pm$ 2.1 & 988 $\pm$ 6 \\
$T_2$ & 38.3 $\pm$ 2.0 & 986 $\pm$ 5 & \textbf{39.9 $\pm$ 2.0} & \textbf{989 $\pm$ 6} & 38.2 $\pm$ 2.6 & 986 $\pm$ 8 & 38.2 $\pm$ 2.1 & 985 $\pm$ 6  \\
FLAIR & 35.4 $\pm$ 2.1 & 978 $\pm$ 9 & 35.7 $\pm$ 1.9 & 978 $\pm$ 7 & \textbf{37.3 $\pm$ 2.1} & \textbf{983 $\pm$ 5} & 35.6 $\pm$ 2.4 & 978 $\pm$ 8  \\
$T_1$-POST & 42.0 $\pm$ 2.5 & 994 $\pm$ 3 & 41.8 $\pm$ 2.0 & 994 $\pm$ 4 & 41.5 $\pm$ 2.0 & 993 $\pm$ 4 & \textbf{43.1 $\pm$ 2.3} & \textbf{995 $\pm$ 3} \\ \hline
\multicolumn{1}{c|}{} & \multicolumn{8}{c|}{MoDL-I-100} \\ \cline{2-9}
\multicolumn{1}{c|}{} & \multicolumn{2}{c|}{$T_1$} & \multicolumn{2}{c|}{$T_2$} & \multicolumn{2}{c|}{FLAIR} & \multicolumn{2}{c|}{$T_1$-POST}\\  
\multicolumn{1}{c|}{} & PSNR & SSIM & PSNR & SSIM & PSNR & SSIM & PSNR & SSIM \\ \hline 
$T_1$ & \textbf{39.9 $\pm$ 1.9} & \textbf{991 $\pm$ 4} & 38.0 $\pm$ 2.3 & 984 $\pm$ 8 & 38.3 $\pm$ 2.5 & 986 $\pm$ 7 & 39.3 $\pm$ 2.0 & 989 $\pm$ 5 \\
$T_2$ & 37.7 $\pm$ 2.6 & 983 $\pm$ 9 & \textbf{39.6 $\pm$ 2.7} & \textbf{989 $\pm$ 5} & 37.6 $\pm$ 2.1 & 983 $\pm$ 7 & 38.0 $\pm$ 2.3 & 984 $\pm$ 6  \\
FLAIR & 35.3 $\pm$ 2.3 & 978 $\pm$ 8 & 35.5 $\pm$ 1.9 & 978 $\pm$ 7 & \textbf{36.9 $\pm$ 2.6} & \textbf{982 $\pm$ 7} & 35.1 $\pm$ 2.7 & 977 $\pm$ 7  \\
$T_1$-POST & 41.3 $\pm$ 2.0 & 993 $\pm$ 3 & 41.2 $\pm$ 1.7 & 993 $\pm$ 5 & 40.9 $\pm$ 2.4 & 992 $\pm$ 5 & \textbf{43.0 $\pm$ 1.9} & \textbf{995 $\pm$ 3}  \\ 
\hline
\end{tabular}
\vspace{1em}
\caption{Within-domain and cross-domain comparisons of Ada-MoDL against MoDL and MoDL-I on fastMRI brain data. The models have been trained and tested on four different contrasts. MoDL-I-6 and MoDL-I-100 represent training with six and 100 subjects, respectively. Ada-MoDL and MoDL are trained with six subjects per acquisition setting. The top section of the table (first four sub-tables) corresponds to the within-domain results, where the models were trained for the acquisition setting of the data. By contrast, the columns in the bottom section (last two sub-tables) correspond to cross-domain comparisons. Here, the columns correspond to the setting for which the models were trained, while the rows correspond to the acquisition setting. We note that the diagonal entries correspond to the cases where the acquisition setting matched the setting for which the models were trained for. Note that the off-diagonal entries are  lower than than the diagonal entries.}
\label{tab:fastmri_comp_tab} 
\end{table}

\section*{Methods}
\label{sec:methods}

\subsection*{Datasets used for validation}
We use two datasets for validation of the proposed framework. The first dataset was acquired by the team and used to determine the optimal parameter settings. Fully sampled brain MR raw data with $T_1$, $T_2$ and FLAIR contrasts were collected from human subjects on an Orian 1.5T and a Galan 3T system (3 x 2 = 6 acquisition settings) using a 16-channel head/neck coil (Canon Medical Systems Corporation, Tochigi, Japan). The human subjects IRB protocol was approved by the local institution, and data were acquired after receiving informed consent from each subject. The results on data from the six acquisition settings are represented as: $T_1$-3T, $T_2$-3T, FLAIR-3T, $T_1$-1.5T, $T_2$-1.5T, and FLAIR-1.5T, respectively. The 2D matrix size was set as 512 x 320 for all the scans. The undersampled raw data were generated retrospectively using masks with 1-D variable density undersampling along the phase-encoding direction. For the experiments, we consider only the poisson disc sampling pattern.  

We use the fastMRI brain \cite{knoll2020fastmri} dataset for our final validation. We use a subset of the fastMRI data consisting of data from 100 training subjects and 20 testing subjects. We use the raw k-space data from $T_1$, $T_2$, FLAIR, and $T_1$-POST contrast acquisitions, collected on 3T scanners (4 x 1 = 4 acquisition settings). The results on data from the four acquisition settings are represented as: $T_1$, $T_2$, FLAIR and $T_1$-POST, respectively. The scans are 2D multi-slice acquisitions with a matrix size of 320 x 320, and the number of slices per acquisition ranges from 12-16. These datasets are retrospectively under-sampled in k-space using 1-D variable density masks undersampling along the phase encoding direction. In this case, too, we consider only the poisson disc sampling pattern.

\subsection*{Network Architecture}
We chose a five-layer CNN shown in Fig. \ref{fig:arch}, for the experiments with the first dataset. Each convolution layer consists of 3 x 3 filters with 64  channels per layer, except the last layer. The five-layer CNN has approximately 113,000 parameters. We used an MLP with four hidden layers, each with 16 features. The size of the MLP's output layer depends on the size of CNN used. In this case, we have feature-scaling weights for 256 features (64 features x 4 layers). The scaling factors generated by MLP modulate the output of the convolution layers. The size of feature-scaling vector $\mathcal F(m)$ depends on the CNN size. The MLP's hidden layer size is chosen through a study shown in Fig. \ref{fig:mlp_size}. The CNN and MLP used for these experiments is shown in Fig. \ref{fig:arch} (c). We compare the Ada-MoDL architecture with MoDL, which uses a denoising CNN of the same size and it is jointly trained with all of the above datasets. We also compare the performance against a MoDL network (MoDL-I), which is trained for each of the above settings independently; this approach needs a separate network for each setting. The number of learnable parameters in the MLP in Ada-MoDL is 5517, which is about 5\% of the parameters of the five-layer CNN. Thus, Ada-MoDL has 5\% additional trainable parameters when compared to MoDL.

For fastMRI experiments, we chose a UNET as the CNN with three pooling/unpooling layers and baseline features of 64. The larger network offered improved performance in the fastMRI setting with more data. The chosen MLP for UNET has four hidden layers of the same size as that of the MLP for the five-layer CNN used for the first dataset. We maintained the same ratio between the number of MLP and UNET features. The size of the output layer of the MLP is chosen to match the number of features of the UNET.

\subsection*{Training and implementation details}
\label{subsec1}

The first dataset is used for multiple purposes, including choosing the size of MLP and studying Ada-MoDL's performance as well as the impact of training data size. The Ada-MoDL is trained with four different acquisition settings ($T_2$-3T, FLAIR-3T, $T_2$-1.5T, $T_1$-1.5T) for the MLP size selection experiments. We consider four different acceleration factors (1.8x, 2.5x, 3.5x, and 4.0x) for retrospective undersampling of the datasets (4 x 4 = 16 acquisition settings). The FLAIR-1.5T data is used to determine the hyperparameters of MLP, while $T_1$-3T data is intentionally left out to study Ada-MoDL's generalizability on an unseen setting.

The Ada-MoDL network with the chosen MLP size is trained with all the acquisition settings except $T_1$-3T, thus resulting in 20 (5 contrast-field strength combinations x 4 acceleration factors) different settings. Ada-MoDL is tested on both seen and unseen acquisition settings for analysis. For the experiments in Table \ref{tab:canon_comp_tab}, the training set consists of five subjects (5 x 1 subject per setting), while inference is performed on six subjects per acquisition setting (total of 30 subjects). To study the impact of the size of the training dataset (Fig. \ref{fig:data_size_plots}), we consider training with S subjects per acquisition setting, with S varying from one to six. In this case, the training dataset consists of up to 30 training subjects (5 x 6), and inference is performed on two subjects per acquisition setting (total of 10 subjects). Similarly, two subjects have been used for inference on the unseen setting $T_1$-3T (Fig. \ref{fig:$T_1$_3t}). 

The acquisition settings are summarized by a five-dimensional conditional vector $m$ from the Ada-MoDL formulation. The first three entries encode the contrast type of the acquisition ($T_1$, $T_2$, and FLAIR) in a one-hot fashion; binary bits are used to indicate which contrast the data belongs to, while the remaining contrast bits are zero. The fourth component is also binary, indicating the field strength (3T or 1.5T). The last entry is a floating point number, which represents the acceleration factor.     

The fastMRI dataset consists of four different acquisition settings, $T_1$, $T_2$, FLAIR, and $T_1$-POST contrasts, with a field strength of 3T. During training, we consider the recovery of the images at four different acceleration factors (1.8x, 2.5x, 3.5x, 4.0x), resulting in a total of 16 different acquisition settings. In this case, we use a five-dimensional conditional vector $m$, whose first four entries encode the contrast of the acquisition ($T_1$, $T_2$, FLAIR, $T_1$-POST) in a one-hot fashion, while the last entry represents the acceleration factor through a floating point number. The trainings are performed using either six or 100 subjects per contrast. The testing is performed using 20 subjects for each kind of contrast.
 
 In both the above training settings, we pre-train the denoiser with a single unrolling step ($K=1$) in \eqref{cnn_meta} and \eqref{dc_meta}; the $\mathcal D$ trained with $K=1$ is used as an initialization to train the network with $K=10$. The same training strategy is used for both MoDL and MoDL-I. All the methods are implemented using PyTorch. The models are optimized using mean-squared error (MSE) loss for 500 epochs with the Adam optimizer \cite{kingma2014adam} at a learning rate of $10^{-4}$. Training and inference are performed on NVIDIA A-100 GPU.

\subsection*{Experiments}

\subsubsection*{Size of the MLP network}
The Ada-MoDL framework adapts the network to different acquisition settings by adapting the scaling weights and the regularization parameter using the MLP. The size of the MLP determines the trade-off between over-fitting and the ability to adapt to new settings, which we study in Fig. \ref{fig:mlp_size}. In the context of the first dataset, we train multiple networks, where the number of features per layer N varies from 2 to 64. We choose the best network based on the performance of Ada-MoDL on the FLAIR-1.5T dataset, a setting that was not used for training. The best ratio between the number of CNN and MLP parameters is chosen for the rest of the experiments. 

\subsubsection*{Comparison against traditional unrolled networks}

We compare the proposed approach against the traditional MoDL scheme, which uses a single network for all the settings. We also consider different MoDL-I networks, trained individually for each different setting. In addition to requiring the storage of multiple models, the performance of MoDL-I might be restricted when the number of available training datasets for each setting is limited. The impact of the size of training data on the reconstruction performance of Ada-MoDL and MoDL-I is studied in Fig. \ref{fig:data_size_plots}. We also study the generalization performance of MoDL and Ada-MoDL on an unseen acquisition setting in Fig. \ref{fig:$T_1$_3t}.

The comparison of the performance of the proposed scheme against the unrolled networks on the first dataset is shown in Table \ref{tab:canon_comp_tab}. Ada-MoDL, trained with one subject for each of the five settings ($T_2$-3T, FLAIR-3T, $T_2$-1.5T, FLAIR-1.5T, $T_1$-1.5T), is compared against MoDL and MoDL-I, both trained with one subject per contrast for the first dataset. The mean PSNR and SSIM values are reported over six subjects for each setting. The qualitative comparisons on 4.0x accelerated datasets are shown in Fig. \ref{fig:$T_2$_flair_3t} and Fig. \ref{fig:$T_1$_$T_2$_flair_15t}. We also compare the proposed approach in the context of the fastMRI brain dataset \cite{knoll2020fastmri} in Table \ref{tab:fastmri_comp_tab}. In the fastMRI setting, we compare the Ada-MoDL scheme trained with the data from four contrasts with six subjects per contrast against MoDL-I, trained for each setting independently using data from 100 subjects (MoDL-I-100). We also include MoDL-I-6 in the study for reference; it is trained with six subjects. The mean PSNR and SSIM values are reported for 20 subjects for each contrast. Qualitative comparisons on fastMRI datasets are shown for 4.0x accelerated data in Fig. \ref{fig:$T_1$_$T_2$_$T_1$p_flair_fmri}. 

\subsubsection*{Evaluation metrics for quantitative comparison}
Reconstructed images are quantitatively evaluated in terms of Structural Similarity Index (SSIM) \cite{wang2004image} and peak signal-to-noise ratio (PSNR), which is given by $\mathbf {PSNR \hspace{2pt} (dB)} = 20.\log_{10} \Big(\frac{{\rm max}(\mathbf x_{gt})}{\|\mathbf x_{r} - \mathbf x_{gt}\|_2}\Big).$ The ${\rm max}(\cdot)$ function provides the maximum pixel value of fully-sampled ground truth $\mathbf x_{gt}$ and $\mathbf x_r$ denotes the reconstructed image.

\section*{Results}
\label{sec:results}

\subsection*{Selection of MLP size/complexity}

\begin{figure}[h!]
\centering
\includegraphics[width=0.5\linewidth,keepaspectratio=true,trim={1.7cm 8.7cm 9.7cm 9.2cm},clip]{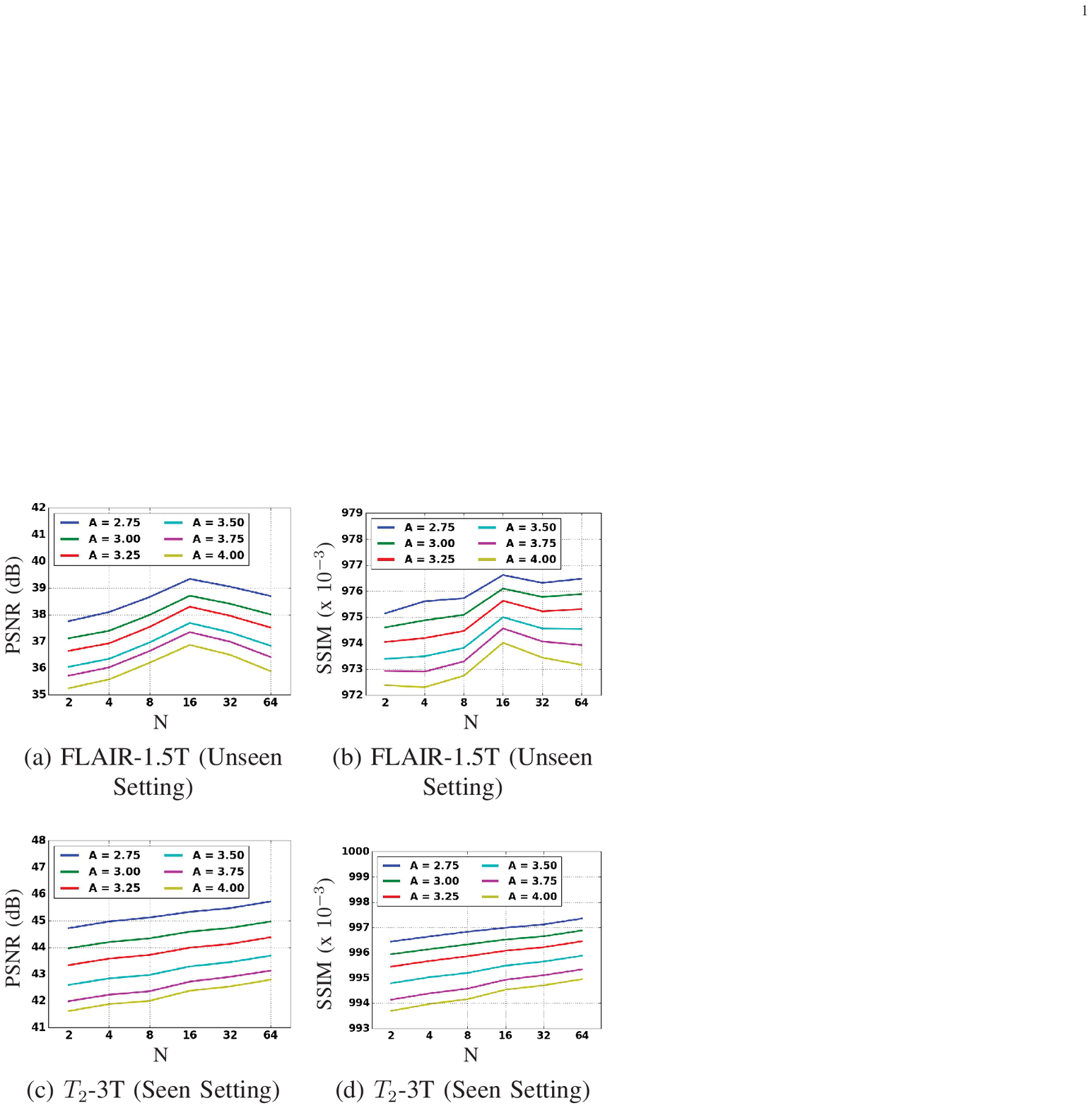}
\caption{Performance comparison of Ada-MoDL with different sizes of MLPs. The graphs are plotted with respect to N={2,4,8,16,32,64,} where N denotes the number of features per layer of the MLP. There are six plots, each corresponding to a different acceleration factor ranging from A = 2.75x to A = 4.0x at an interval of 0.25x. The models have been tested on FLAIR-1.5T and $T_2$-3T datasets. $T_2$-3T is one of the acquisition settings from the training set while FLAIR-1.5T is an unseen setting during training. The first row shows PSNR and SSIM plots ((a), (b)) for FLAIR-1.5T, while the second row represents those ((c), (d)) for $T_2$-3T. Trends show that higher N provides better performance on a seen setting from the training set; on the other hand, performance reduces for $\rm N > 16$ when tested on an unseen setting.}
\label{fig:mlp_size}
\end{figure}

The comparison of Ada-MoDL, trained and tested on the first dataset, with different sizes of MLP is shown in Fig. \ref{fig:mlp_size}. In particular, we show the quality of the reconstructions at different acceleration factors for six models, which differ in the number of features per layer of the MLP denoted by N chosen from N = \{2,4,8,16,32,64\}. The PSNR (dB)/SSIM versus N plots are shown for six different acceleration factors ranging from A=2.75x to 4.0x with an interval of 0.25x.    
The top row shows the performance on the FLAIR-1.5T acquisition setting, which was not used for training, while the bottom row corresponds to the $T_2$-3T setting which was included in training.
It is seen that, for the $T_2$-3T setting, the performance improves consistently with an increase in size of the MLP. The MLP with the largest size offers the best performance on $T_2$-3T at all acceleration factors, as expected. By contrast, the MLP with N=16 features per layer offers improved performance when compared to other models on the FLAIR-1.5T dataset; this is visible through a peak at N=16. The performance drops with further increase in MLP size. This trend on FLAIR-1.5T is observed at all the acceleration factors. These results show that N=16 offers a better generalization performance. As noted before, the number of parameters of this MLP is around 5\% of that of the CNN module.

\begin{figure}[h!]%
\centering
\includegraphics[scale=0.95,keepaspectratio=true,trim={1.8cm 10.2cm 13.4cm 10.7cm},clip]{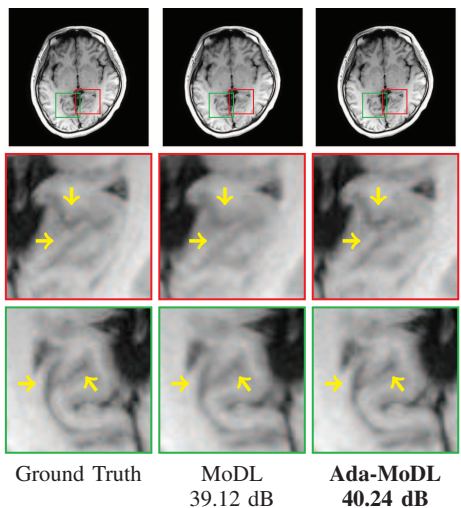}
\caption{Generalization performance: Reconstruction results on 4.0x accelerated brain data from the $T_1$-3T setting. The images shown above are magnitude images showing differences between Ada-MoDL and MoDL in terms of reconstruction quality for $T_1$-3T datasets. Note that $T_1$-3T is an unseen setting for the models during training. We observe reduced blurring, compared to the MoDL seting, demonstrating the ability of the Ada-MoDL network to generalize to an unseen setting.}%
\label{fig:$T_1$_3t}%
\end{figure}

In Fig. \ref{fig:$T_1$_3t}, we compare the performance of Ada-MoDL with the best MLP size against MoDL on the $T_1$-3T data at 4x acceleration, which is an acquisition setting not seen by MoDL or Ada-MoDL during training. The zoomed red and green regions show significant blurring with MoDL, while the Ada-MoDL scheme is able to offer sharper reconstructions. In terms of PSNR, Ada-MoDL offers more than 1 dB improvement over MoDL. The improved performance offered by Ada-MoDL can be explained by the ability of the MLP to extrapolate the scaling weights and the regularization parameters from the acquisition settings it has seen to an unseen acquisition setting. The experiments in Fig. \ref{fig:$T_2$_flair_3t} and Fig. \ref{fig:$T_1$_$T_2$_flair_15t} correspond to datasets from 3T and 1.5T, respectively, which are acquisition settings that were used in training. These results show that the above choice of the MLP size also translated to good performance in these cases.

\begin{figure}[t!]
\centering
\includegraphics[width=0.5\linewidth,keepaspectratio=true,trim={1.8cm 11.5cm 9.7cm 11.9cm},clip]{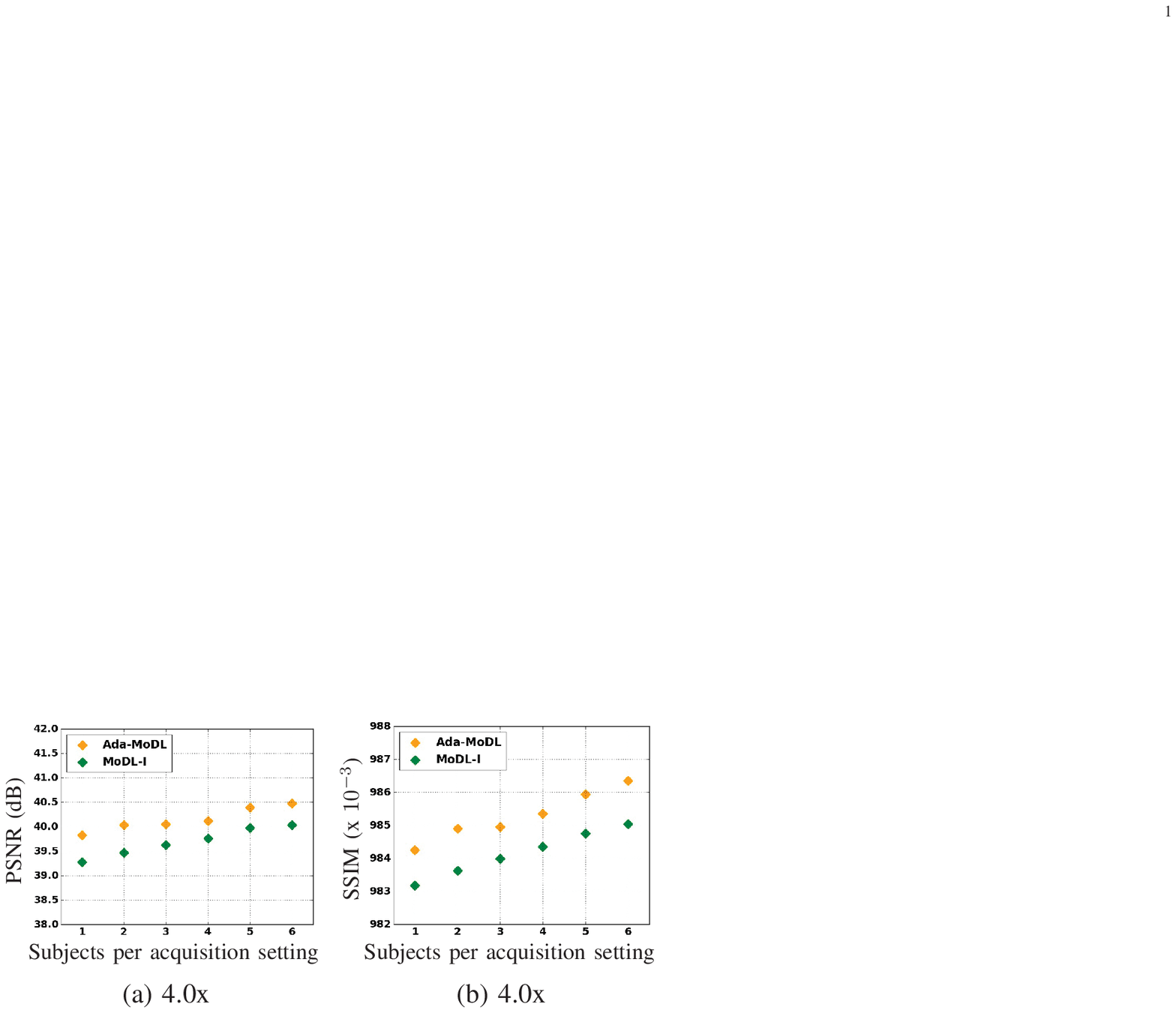}
\caption{Impact of the number of training subjects on performance. (a) indicates the plot of the PSNR as a function of subjects, while (b) shows the plot of SSIM. Ada-MoDL is compared against MoDL-I on 4.0x accelerated $T_1$-1.5T data. Ada-MoDL has been trained on multiple acquisition settings with S subjects from each setting, while MoDL-I has been trained only on $T_1$-1.5T with S subjects from it. The models are trained for S = 1,2,3,4,5, and 6. The average PSNR and SSIM plots show how the image quality changes with the number of training subjects. The results show that the performance of Ada-MoDL for one subject per acquisition setting is comparable to what MoDL-I can give with four subjects.}
\label{fig:data_size_plots}
\end{figure}

\subsection*{Impact of training dataset size}
The MoDL-I network, where a model is trained with data from only one acquisition setting, offers the best performance when sufficient training data from that setting is available. However,, the performance of this approach is limited in practical scenarios, when insufficient training data is available for each specific acquisition setting. In addition, this approach require the storage of multiple trained models on the scanner. The performance of Ada-MoDL and MoDL-I on test data from $T_1$-1.5T is plotted against the number of training subjects S per acquisition setting, ranging from one to six in Fig. \ref{fig:data_size_plots}. While Ada-MoDL has been trained on five acquisition settings with four different acceleration factors, leading to 20 different settings, MoDL-I has been trained on only 4.0x accelerated $T_1$-1.5T datasets (one setting). We note that the $T_1$-1.5T is a setting that was used during training for both models. In both plots, it is observed that the performance gap narrows with training subjects increasing from one to four, while it remains steady from four to six. MoDL-I shows greater improvement with an increase in data, and it is expected to perform on par with or better than Ada-MoDL with sufficient availability of training subjects. It is evident from the plot that Ada-MoDL trained with fewer subjects per setting (one) is performing on par with MoDL-I trained with higher number of subjects (four) from that setting. The ability of Ada-MoDL to learn from datasets acquired with different acquisition settings translates to improved performance over MoDL-I. This implies that fewer datasets are needed for each setting to train the Ada-MoDL network. We also do a similar comparison using fastMRI dataset in the next section; we compare the performance of Ada-MoDL trained with six subjects per setting against that of MoDL-I trained with 100 subjects.

\subsection*{Quantitative evaluation on the first dataset}
We compare Ada-MoDL against MoDL and MoDL-I on 20 different acquisition settings (three contrasts, two field strengths and four acceleration factors) in the context of the first (acquired in-house) dataset in Table \ref{tab:canon_comp_tab}. We train all the networks with one subject per acquisition setting. The inference results on testing subjects with data acquisition settings (contrast, field strength, acceleration) are similar to the ones used for training are shown in Table \ref{tab:canon_comp_tab}. The table consists of five sub-tables, corresponding to FLAIR-3T, $T_2$-3T, FLAIR-1.5T, $T_2$-1.5T, and $T_1$-1.5T. The sub-tables report PSNR and SSIM values at different acceleration factors (1.8x, 2.5x, 3.5x, 4.0x). It is observed that at lower acceleration factors (e.g., 1.8x, 2.5x), the gap in performance between Ada-MoDL and MoDL-I is small. However, at higher acceleration factors (e.g., 3.5x and 4.0x), Ada-MoDL offers improved performance when compared to MoDL-I. In particular, the ability of Ada-MoDL to combine information from multiple types of datasets translates to improved performance, especially when few training datasets are available for each type of acquisition. The performance of MoDL-I will improve with more training datasets, as shown in Fig. \ref{fig:data_size_plots}. The experiments in this figure show the ability of Ada-MoDL to offer good performance, even when few training  datasets per acquisition setting is available. Ada-MoDL has been compared pairwise with MoDL and MoDL-I using the Wilcoxon signed-rank test. It was found to be statistically significant with a p-value of $\rm p < 0.05$.

We note that the performance of the MoDL scheme is consistently lower than that of the other methods. Note that the MoDL network weights as well as the regularization parameter are the same for all the acquisition settings; the average parameter values might not be optimal for any of the acquisition settings, translating to lower performance.

\begin{figure}[t!]%
\centering
\includegraphics[scale=0.9,keepaspectratio=true,trim={1.5cm 8.2cm 9.6cm 9cm},clip]{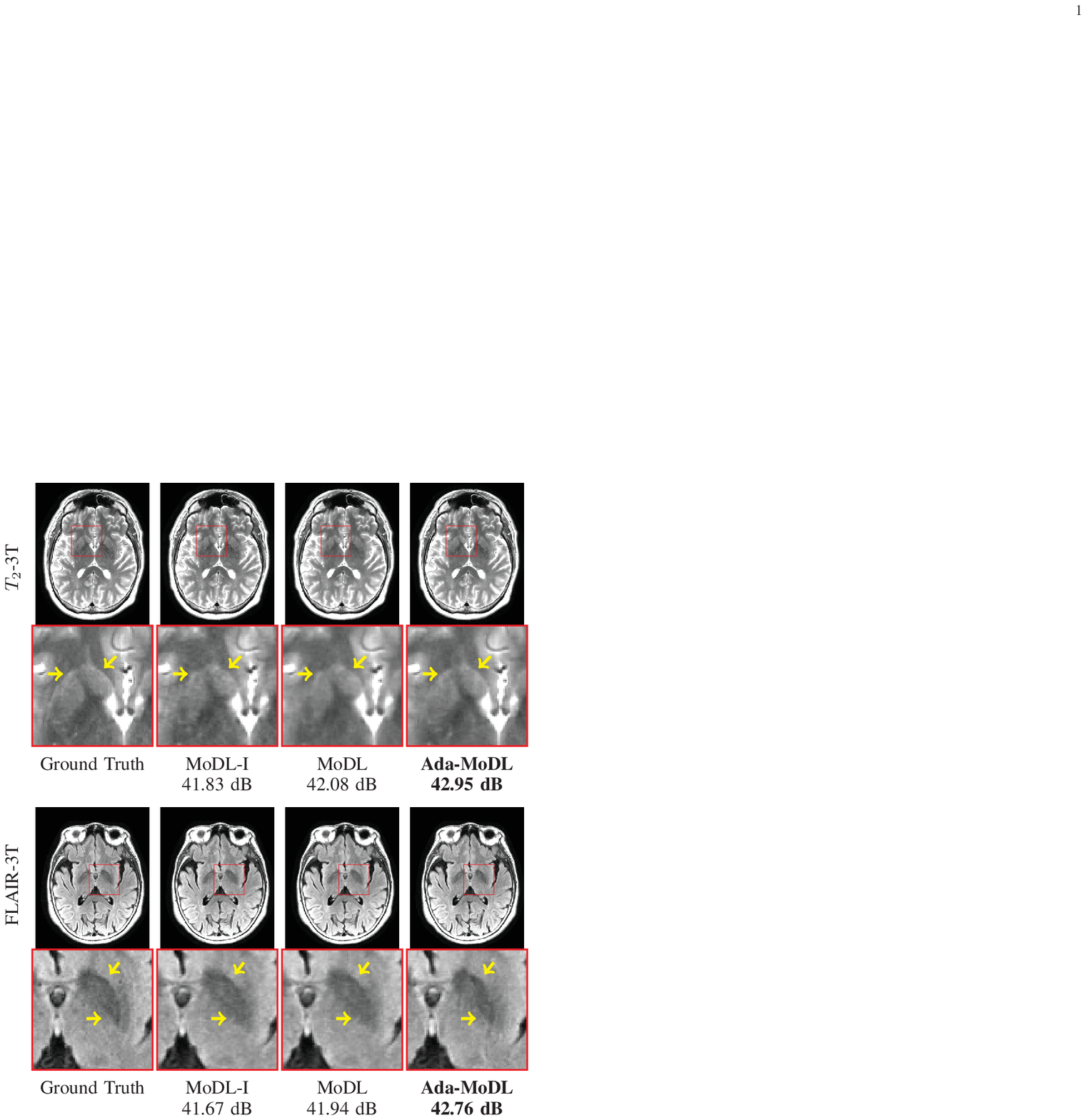}
\caption{Reconstruction results on 4.0x accelerated brain data from 3T. The images shown above are magnitude images showing differences between Ada-MoDL, MoDL, and MoDL-I in terms of reconstruction quality for $T_2$-3T (top) and FLAIR-3T datasets (bottom). The arrows in the zoomed images indicate regions with differences in image details. MoDL-I is trained with one subject, while Ada-MoDL and MoDL are trained with one subject per acquisition setting. The quantitative results averaged over testing subjects are shown in Table \ref{tab:canon_comp_tab}.}%
\label{fig:$T_2$_flair_3t}%
\end{figure}

\begin{figure}[t!]%
\centering
\includegraphics[scale=0.95,keepaspectratio=true,trim={1.6cm 6.2cm 11.1cm 7.1cm},clip]{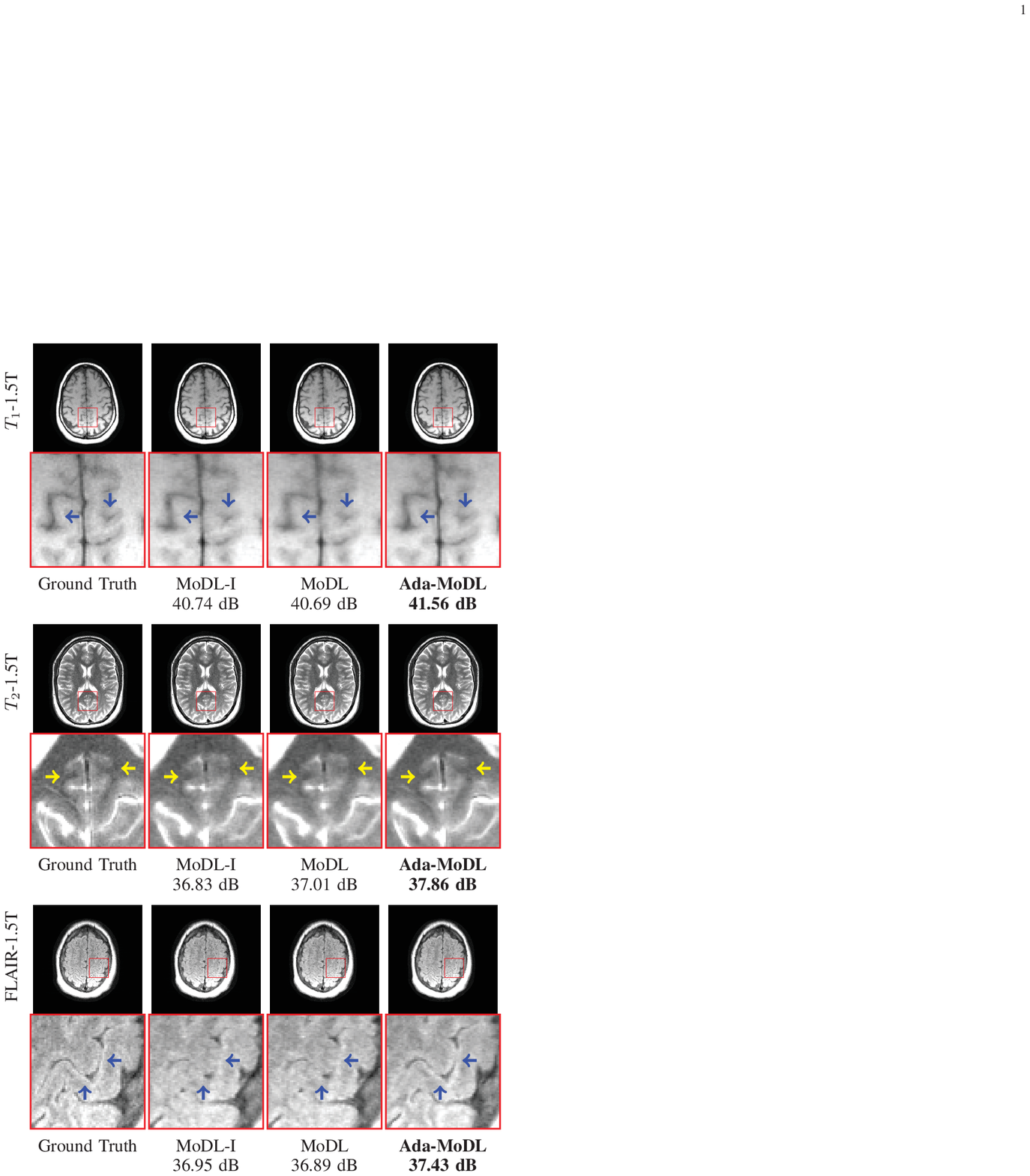}
\caption{Reconstruction results on 4.0x accelerated brain data from 1.5T. The images shown above are magnitude images showing differences between Ada-MoDL, MoDL, and MoDL-I in terms of reconstruction quality for $T_1$-1.5T, $T_2$-1.5T, and FLAIR-1.5T datasets. The arrows in the zoomed images indicate regions with differences in image details. MoDL-I is trained with one subject, while Ada-MoDL and MoDL are trained with one subject per acquisition setting. The quantitative results averaged over testing subjects are shown in Table \ref{tab:canon_comp_tab}.}%
    \label{fig:$T_1$_$T_2$_flair_15t}%
\end{figure}

Visual comparisons of the Ada-MoDL reconstructions of 3T datasets at 4.0x acceleration with those of MoDL and MoDL-I are shown in Fig. \ref{fig:$T_2$_flair_3t} and Fig. \ref{fig:$T_1$_$T_2$_flair_15t}, respectively. In Fig. \ref{fig:$T_2$_flair_3t}, the arrows point towards the regions with differences in reconstruction. For $T_2$-3T, the zoomed basal ganglia of the brain shows sharper edges in the Ada-MoDL reconstruction. The MoDL reconstruction appears blurred, while MoDL-I looks more noisy. In terms of SNR, Ada-MoDL performs $\approx$ 1 dB better than the other two methods. A similar observation is noted in a FLAIR-3T slice, with Ada-MoDL showing sharper edges and lower noise. As indicated by arrows, both MoDL and MoDL-I appear blurred. 

Figure \ref{fig:$T_1$_$T_2$_flair_15t} shows the reconstructed images on 1.5T; the rows correspond to $T_1$-1.5T, $T_2$-1.5T, and FLAIR-1.5T scans, respectively. For FLAIR-1.5T, arrows indicate better preservation of details in Ada-MoDL, which appears to be lost in the other two. In $T_1$-1.5T, the blue arrows indicate relatively sharper edges for Ada-MoDL, and a similar trend is observed for $T_2$-1.5T. Ada-MoDL also seems to preserve details more accurately than MoDL and MoDL-I. In terms of SNR, Ada-MoDL offers  an improved margin of 0.6-0.8 dB. Note that MoDL learns a single $\lambda$ for all the contrasts and hence suffers from lack of flexibility in balancing the effects of DC and prior term. MoDL-I requires several models since it requires a model to be trained for every acquisition setting, including contrast, acceleration, field strength, etc., which is prohibitive. Ada-MoDL addresses these limitations with a conditional framework, as shown in these performance comparisons.

\begin{figure}[t!]%
\centering
\includegraphics[scale=0.95,keepaspectratio=true,trim={1.6cm 3.6cm 11.1cm 4.4cm},clip]{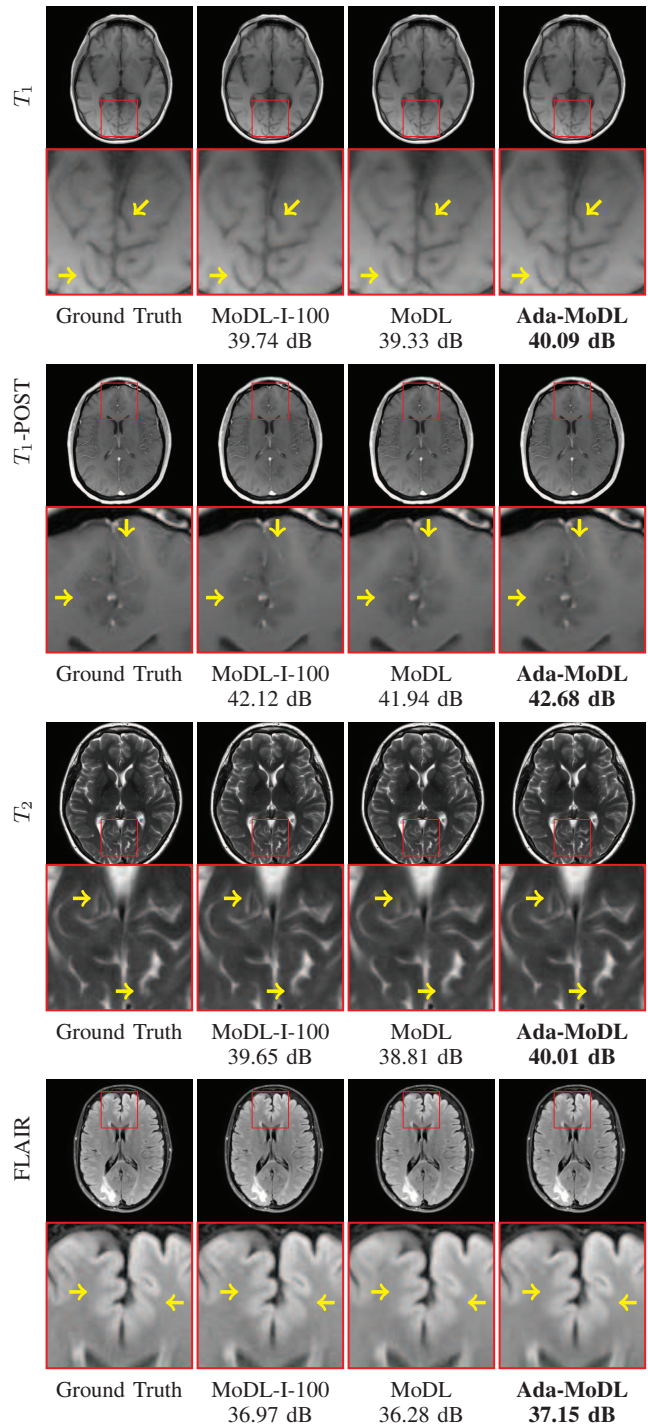}
\caption{Reconstruction results on 4.0x accelerated fastMRI brain data. The images shown above are magnitude images showing differences between Ada-MoDL, MoDL, and MoDL-I-100 in terms of reconstruction quality for $T_1$, $T_2$, $T_1$-POST, and FLAIR datasets. MoDl-I-100 denotes MoDL-I trained with 100 subjects. Ada-MoDL and MoDL are trained with six subjects per acquisition setting. The quantitative results averaged over testing subjects are shown in Table \ref{tab:fastmri_comp_tab}.}%
    \label{fig:$T_1$_$T_2$_$T_1$p_flair_fmri}%
\end{figure}

\begin{figure}[t!]
\centering
\includegraphics[width=0.5\linewidth,keepaspectratio=true,trim={1.7cm 6.5cm 9.7cm 6.8cm},clip]{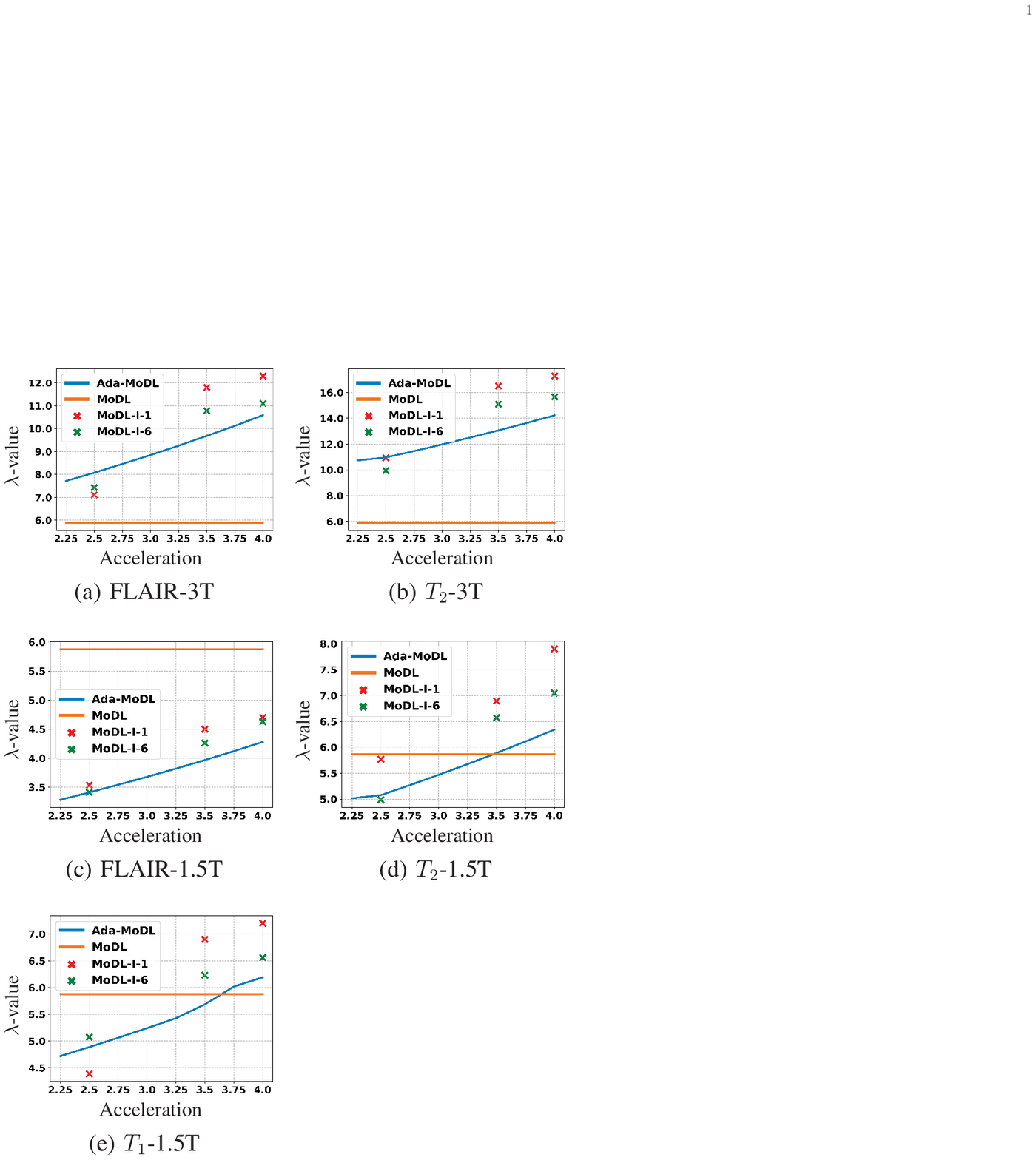}
\caption{Plot of $\lambda$ for Ada-MoDL, MoDL, and MoDL-I with respect to acceleration factors ranging from 2.25 to 4.0 at an interval of 0.25. We show values for five different acquisition settings (FLAIR-1.5T, $T_2$-1.5T, $T_1$-1.5T, FLAIR-3T, $T_2$-3T) from the data collected by the team. MoDL-I-1 and MoDL-I-6 correspond to training with one and six subjects, respectively. Ada-MoDL and MoDL are trained with one subject per setting.}
\label{fig:lambda_plots}
\end{figure}

\makeatletter 
\renewcommand{\thefigure}{S\@arabic\c@figure}
\makeatother
\setcounter{figure}{0}

\begin{figure}[t!]%
\centering
\includegraphics[scale=0.93,keepaspectratio=true,trim={1.6cm 3.6cm 9.0cm 4.4cm},clip]{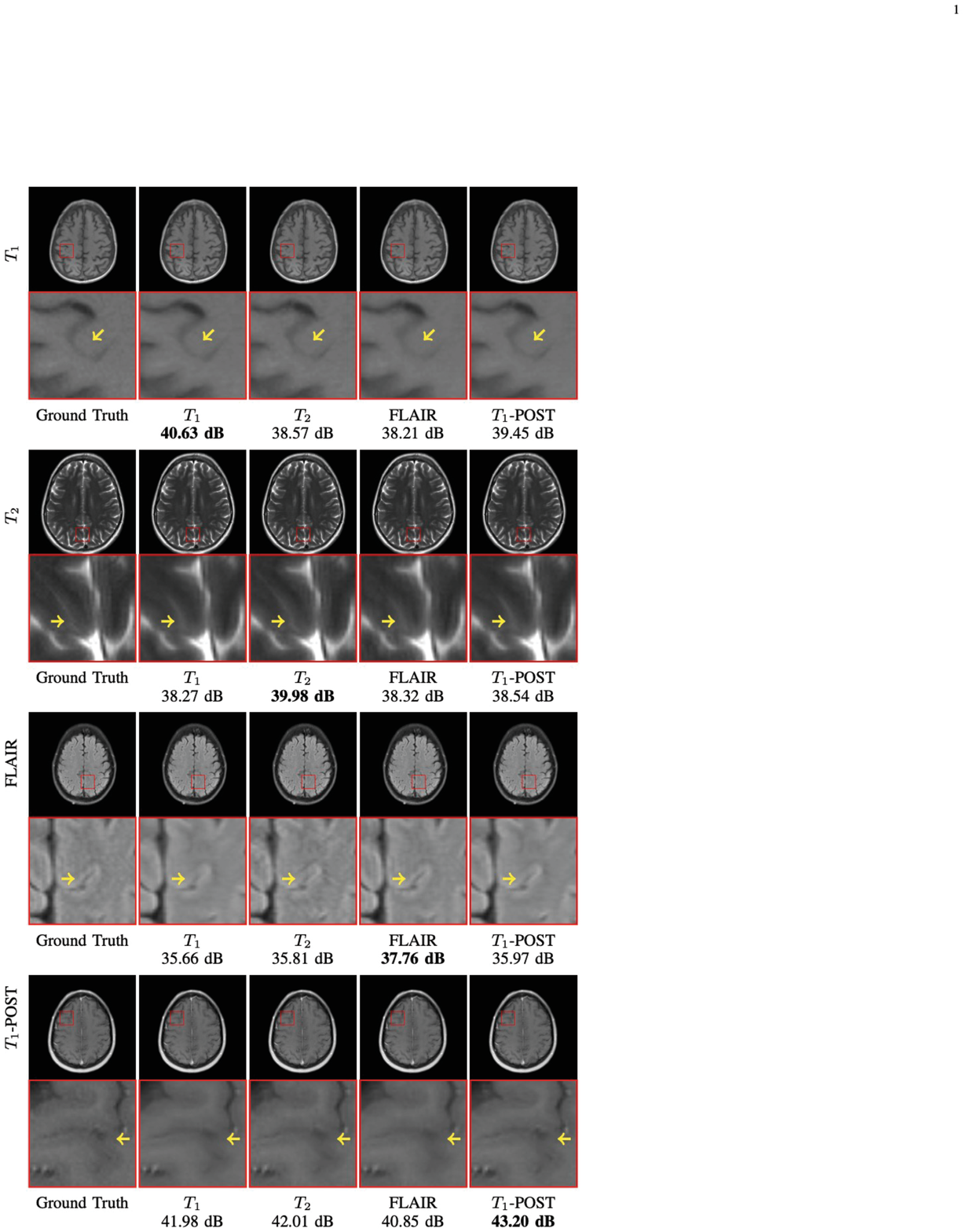}
\caption{Cross-Domain reconstruction results for Ada-MoDL on 4.0x accelerated fastMRI brain data. The rows correspond to the acquisition setting of the image and the columns denote the setting for which the network parameters are generated using the MLP. The within-domain performances along the diagonal are the best in each case (highlighted in bold) which is expected due to the network parameters that are matched to the setting while cross-domain performances drop due to mismatch. These are zoomed magnitude images with arrows pointing out the differences. The quantitative results averaged over testing subjects are shown in Table 2 in the main paper.}%
    \label{fig:$T_1$_$T_2$_$T_1$p_flair_fmri_cd}%
\end{figure}

\subsection*{Quantitative evaluation on fastMRI Brain dataset \cite{knoll2020fastmri}}

\subsubsection*{Performance on Within-Domain Data}

We ran experiments on fastMRI brain data \cite{knoll2020fastmri} to compare the performance of Ada-MoDL against that of MoDL and MoDL-I. Ada-MoDL and MoDL are trained with six subjects per contrast. We consider two versions of MoDL-I, namely, MoDL-I-6 and MoDL-I-100. MoDL-I-6 is trained with six while MoDL-I-100 with 100 subjects. The models have been tested on 20 subjects for each contrast, including $T_1$, $T_2$, FLAIR, and $T_1$-POST. The inference results on test datasets are shown through the first four sub-tables under the category "Within-Domain Data" in Table \ref{tab:fastmri_comp_tab}. Each sub-table corresponds to a specific contrast with four different acceleration factors (1.8x, 2.5x, 3.5x, 4.0x). The within-domain result corresponds to inference on an acquisition setting using a model with learned parameters for the same acquisition setting. 

The performance of MoDL-I improves upon increasing training subjects from six to 100 ($\approx 1$ dB for $T_1$, $T_2$, FLAIR, and $\approx 0.5$ dB for $T_1$-POST). Ada-MoDL outperforms MoDL and MoDL-I-6 while performing slightly better than MoDL-I-100. The improved performance of Ada-MoDL shows the benefit of using an MLP-based network for combining information from multiple contrasts. Although MoDL is trained with data from all the contrasts, it learns only one set of parameters for all of them, which provides sub-optimal performance compared to Ada-MoDL. The differences in Ada-MoDL's performance compared to other methods were measured through the Wilcoxon's signed rank test, which gave a significance level of $\rm p < 0.05$.

Reconstruction results of the above-mentioned methods are shown in Fig. \ref{fig:$T_1$_$T_2$_$T_1$p_flair_fmri}. The images are reconstructed from 4.0x under-sampled data. The rows correspond to reconstructions for $T_1$, $T_2$, FLAIR, and $T_1$-POST, respectively. Overall, the performance of MoDL is inferior to MoDL-I-100 in terms of PSNR and differences in reconstruction quality as indicated by yellow arrows. On the other hand, Ada-MoDL performs on par with or slightly better than MoDL-I-100. The arrows point to the regions of the brain where features are much sharper for Ada-MoDL and MoDL-I-100 than for MoDL. This result suggests that Ada-MoDL trained with as few as six subjects per contrast can offer performance equivalent to that of MoDL-I trained with 100 subjects and, therefore, is efficient in terms of training data requirement. Since MoDL's parameters are optimal for none of them, it provides relatively poor performance.

\subsubsection*{Performance on Cross-Domain Data}
We also consider inference using parameters learned for a different acquisition setting. These results are reported in the bottom two sub-tables under the category "Cross-Domain Data" in Table \ref{tab:fastmri_comp_tab}. One sub-table corresponds to results of Ada-MoDL, while the other corresponds to those of MoDL-I-100. Ada-MoDL has one model; the columns represent acquisition settings for which the learned parameters are chosen through the MLP, while the rows represent the settings on which they are tested. Since MoDL-I-100 has different models for different acquisition settings, each column represents a model trained for the specific acquisition setting, and the row corresponds to the setting it is tested on. This performance study has been conducted only on the settings used for training.

The diagonal entries (within-domain) in the table are the bold ones providing better performance where the chosen models/parameters are matched with the settings. Ada-MoDL's performance drops $\approx 1.5-2.0$ dB in cross-domain settings, which is evident from the off-diagonal entries. For MoDL-I-100, the performance degradation is slightly more $\approx 2.0-2.5$ dB in the cross-domain settings. We note that MoDL-I-100 learns a distinct model for each of the acquisition settings and, therefore, the drop is relatively more due to a greater mismatch in the parameters, as expected. On the other hand, Ada-MoDL has same the CNN parameters across settings except for the scaling factors and regularization parameters, which are generated conditionally by the MLP. The reconstruction results are shown in Fig. S1 in the supplementary material. The cross-domain reconstructions are lower in PSNR and also miss out on some details as indicated by the arrows; those details are preserved in within-domain reconstructions (diagonal images).

\subsection*{Adaptation of the network to different settings}
We study the trend of variation in the regularization parameter $\lambda$ with respect to acceleration factors for multiple acquisition settings in Fig. \ref{fig:lambda_plots}. The $\lambda$ values for MoDL, Ada-MoDL, and MoDL-I are plotted for acceleration factors ranging from 2.25x to 4.0x. We note that MoDL uses a single $\lambda$ of value $\approx6.5$ for all acquisition configurations. By contrast, the regularization parameter $\lambda$ in Ada-MoDL is derived using an MLP as a function of meta-data, which varies with the acquisition settings. MoDL-I learns an optimal $\lambda$ for a specific acquisition setting. 

The plots in Fig. \ref{fig:lambda_plots} demonstrate how the proposed approach adapts the network to different settings. The parameter $\lambda$ controls the tradeoff between DC and the regularization term. As the acceleration increases, fewer data samples are available, making the DC term $\|\mathcal A(\mathbf x) - \mathbf b\|_2^2$ smaller in magnitude. To compensate, the $\lambda$ that weighs the DC needs to be increased to offer the best performance. The experiments also show that the regularization parameter values for the 3T datasets are higher than those of the 1.5T datasets. Note that the 3T datasets have higher SNR, and hence the data can be trusted more; a higher regularization parameter adds this prior information to the reconstruction network. The results also show that the Ada-MoDL $\lambda$ values are closer to those of MoDL-I-6 than those of MoDL-I-1. Note that both Ada-MoDL and MoDL-I-6 offer improved performance when compared to MoDL and MoDL-I-1.

\section*{Discussion}
\label{sec:discussion}

Ada-MoDL has been compared against MoDL-I, which is trained for a specific setting and also against MoDL, which learns a single network for all settings. The experiments performed on the smaller dataset have been used to determine MLP size and study performance; based on those findings, the method has been validated on the larger dataset from fastMRI. We observe that the performance of the Ada-MoDL scheme, which can adapt the representation to each specific setting, offers better performance than the non-adaptive MoDL in all cases, irrespective of dataset size. The gap in performance increases with the size of the dataset. The lower generalizability of MoDL can be attributed to its learning a compromise choice of parameters for multiple settings, which is sub-optimal for each of the settings considered. We note that MoDL-I learns a specific set of CNN filters for each setting; this approach offers improved performance than MoDL. By contrast Ada-MoDL learns a single set of CNN filters from all the different settings, while modulating the feature weights and the regularization parameter for the unrolled network to adapt to a specific acquisition setting. The MLP learns to weight specific features depending on the acquisition settings, which translates to better performance. Similarly, the  regularization parameter $\lambda$ is expected to vary depending on the signal to noise ratio of the measurements and how undersampled the acquisition is.  Ada-MoDL's setting-dependent parameter $\lambda(m)$ obtained through the MLP enables the adaptation of the suitable $\lambda$ for each setting, thus improving performance. MoDL-I is trained for only one acquisition setting and hence overfits when there are fewer datasets available from that setting; this explains its lower performance in Table \ref{tab:canon_comp_tab}.

Ada-MoDL's performance is validated on a larger brain dataset from the fastMRI \cite{knoll2020fastmri}. It is compared against MoDL and MoDL-I for within-domain as well as cross-domain data. On within-domain data, Ada-MoDL outperformed MoDL and MoDL-I-6 significantly while performing on par with MoDL-I-100. It is noteworthy that Ada-MoDL, trained with six subjects per acquisition setting, could perform on par with MoDL-I-100 (trained on 100 subjects). Thus, Ada-MoDL requires relatively fewer training datasets. Ada-MoDL's performance drops $\approx 2$ dB on domain-shifts, which shows that the parameters for a specific setting do not work well for a different one. A similar behavior is observed for MoDL-I-100, as shown in Table \ref{tab:fastmri_comp_tab}. This implies that the MLP for Ada-MoDL is able to provide very distinct feature weights as well as regularization parameter $\lambda$ for each setting.

The Ada-MoDL approach makes it easier to deploy deep unrolled networks in a clinical setting; rather than using multiple networks for each specific acquisition setting, the parameters can be adapted depending on the acquisition settings. In this work, we only considered the dependence of the model parameters on acceleration, field strength, and contrast. We used only Poisson disc sampling density for our experiments. Ada-MoDL can incorporate multiple parametric sampling densities. Specifically, the parameters of the density (e.g., standard deviation of the distribution) can be used as one of the conditional vectors, while the type of density (e.g, Gaussian, Uniform) may be encoded by one-hot encoding. However, these detailed studies are beyond the scope of this work. An extension of Ada-MoDL to more variations, including anatomy, signal to noise ratio, coil arrays, field of view (FOV), matrix sizes, etc., would warrant defining the conditional vector $m$ differently. While we choose to use one-hot encoding for including variations in acquisitions not considered in our experiments, one may add FOV and matrix sizes as integer conditional vectors instead. Thus, there can be multiple ways to define $m$, which could affect its structure of the condition vector. Our experiments show the ability of the network to adapt to a setting that was not included in the original training. 

A challenge with our current implementation is that the network need to be retrained, when the definition of the condition vector changes. In this setting, one may keep the CNN weights that is learned from a large enough set of acquisition conditions frozen, while only the MLP parameters may be retrained to the new setting; the CNN filters that are learned from multiple settings may be sufficient for the new setting. However, this experiment is beyond the scope of the present work.

The performance improvement offered by Ada-MoDL is higher in Figure \ref{fig:$T_1$_$T_2$_flair_15t}. We note that the fully sampled reference data is noisy in the 1.5T setting, especially for the FLAIR constrast. The MoDL-I approach suffers from the poor quality of the training data in this setting. The improved performance of Ada-MoDL may be attributed to its ability to transfer the learned image features from other acquisition settings to the low SNR settings.

In this work, we only considered the adaptation of the unrolled MoDL algorithm. However, the proposed approach is general enough to be applied to all unrolled algorithms, including \cite{lecun2015deep,hammernik2018learning,aggarwal2018modl,pramanik2020deep, hammernik2022machine, hammernik2022physics, adler2018learned, hammernik2020sigma, liang2020deep}. The proposed concept can also be extended to any conditional deep learning network that includes direct inversion schemes, such as UNET \cite{han2019k}. The CNNs involved in these approaches can be made adaptive thorugh an MLP in a similar fashion as that proposed for Ada-MoDL.

\section*{Conclusion}
\label{sec:conclusion}

We propose a conditional unrolled DL approach called Ada-MoDL for parallel MRI recovery. Unlike traditional unrolled DL methods, the proposed approach has learnable parameters that are functions of acquisition information of the dataset to be recovered. It provides a single network for image recovery from multiple acquisition settings, including contrasts, field strengths, and acceleration factors. The ability of Ada-MoDL to adapt to the acquisition condition translates to improved performance over traditional unrolled methods. The joint training of the Ada-MoDL network on multiple datasets was seen to be more training-data efficient than training individual unrolled networks for each specific acquisition condition. The results also show that Ada-MoDL is able to extrapolate to acquisition settings that are not seen during training. The lightweight architecture is thus expected to make the deployment of unrolled architecture in clinical settings more efficient.

\section*{Acknowledgement}
\label{sec:acknowledgement}
The authors thank Kensuke Shinoda at Canon Medical Systems Corporation for sharing a dataset that was used in this work.

\bibliography{main}

\end{document}